\documentclass[12pt]{iopart}
\usepackage{iopams}
  
\usepackage{graphicx}
\graphicspath{{Figures/}}
\usepackage[colorlinks=true]{hyperref}
\usepackage[numbers,sort&compress]{natbib}

\newcommand{\YR}{YbRu$_2$Ge$_2$}
\newcommand{\CR}{CeRh$_2$As$_2$}
\newcommand{\CB}{CeB$_6$}

\makeatletter
\@ifundefined{text}{%
  \def\text#1{%
    \relax
    \ifmmode
      \mathchoice
        {\hbox{{\everymath{\displaystyle     }#1}}}%
        {\hbox{{\everymath{\textstyle        }#1}}}%
        {\hbox{{\everymath{\scriptstyle      }\let\f@size\sf@size\selectfont#1}}}%
        {\hbox{{\everymath{\scriptscriptstyle}\let\f@size\ssf@size\selectfont#1}}}%
      \glb@settings
    \else
      \mbox{#1}%
    \fi
  }
}{}
\makeatother

\begin{document}

\title[{\footnotesize Thermodynamics, elastic anomalies and excitations in the 
field induced phases of CeRh$_2$As$_2$}]{Thermodynamics, elastic anomalies and excitations in the 
field induced phases of CeRh$_2$As$_2$}

\author{Peter Thalmeier$^{(1)}$, Alireza Akbari$^{(2)}$ and Burkhard Schmidt$^{(1)}$}
\address{$^{1}$Max Planck Institute for the  Chemical Physics of Solids, D-01187 Dresden, Germany}
\address{$^{2}$Beijing Institute of Mathematical Sciences and Applications (BIMSA), Huairou District, Beijing 101408, China}
\ead{bs@cpfs.mpg.de}
\date{\today}

\begin{abstract}
The tetragonal heavy fermion compound \CR~exhibits  unconventional superconductivity 
accompanied by other broken symmetry phases that have been identified as presumably
small moment  intrinsic antiferromagnetism at low magnetic fields and induced quadrupolar order at higher in-plane  fields.
The latter may extend to very large pulsed-field range. The phase boundaries can be investigated by 
following thermodynamic  anomalies like specific heat, magnetocaloric coefficient, thermal expansion and magnetostriction. We calculate their discontinuities and identify the influence of the field induced quadrupole on them. Furthermore we investigate the elastic constant anomalies which are determined by the static homogeneous quadrupolar RPA response functions. We present a calculation of these anomalies for the appropriate symmetry mode both in the disordered and ordered
regime and investigate their change with applied field. In addition we consider the dynamical momentum dependent magnetic susceptibility and the associated dispersion of low energy magnetic modes and how their characteristics change across the phase boundary.
\end{abstract}

\maketitle

\section{Introduction}
\label{sec:intro}

The H-T phase diagram of the tetragonal (space group $D^7_{4h}$) heavy fermion superconductor \CR~has been investigated by several experimental methods \cite{khim:21,hafner:22,mishra:22,semeniuk:23,chajewski:24,khanenko:25} and found to be extremely anisotropic with respect to the field direction. This applies to the superconducting as well as the normal state broken symmetry phases. For field along tetragonal c-axis there are two superconducting phases observed which have been proposed to belong to different (e.g. pseudo-spin singlet or triplet) irreducible representations. This possibility has been attributed to the locally noncentrosymmetric structure  where Ce-4f electrons occupy equivalent sites (forming a body centered tetragonal Ce Bravais lattice) with point group $C_{4v}$ that are not inversion centers \cite{landaeta:22,ogata:23,fischer:23,semeniuk:24} although overall inversion symmetry is preserved. Likewise the microscopic nature of the order parameters in the broken symmetry phases of the normal state  is not yet known with certainty. The anisotropy of the phase diagram is inverted in this case where two normal-state phases occur for field perpendicular to the tetragonal axis. Recent $\mu$SR experiments \cite{khim:25} suggest that the normal state low field phase below $T_0=0.5$ K may be of magnetic nature with presumably
small moments and quasistatic character. Ferromagnetism is excluded due to lack of thermodynamic evidence and antiferromagnetism with small moments seems a possibility. This is supported by correlated band structure calculations \cite{wu:24} that show the existence of a nesting vector ${\bf q}=(\pi,\pi)$ with a corresponding maximum in the staggered susceptibility at {\bf q}. Indeed inelastic neutron scattering (INS) has demonstrated the existence of spin fluctuation maxima at those commensurate peaks \cite{chen:24a} suggesting an antiferromagnetic structure with ordering vector $(\pi,\pi)$ and staggered moments along the tetragonal axes corresponding to two magnetic sublattices (Fig.\ref{fig:struc}).

However, the anisotropy of the phase boundary, in particular the appearance of a second high field phase for in-plane field is not compatible with simple AF order. Therefore it was suggested in Ref.~\cite{schmidt:24} that for fields in the tetragonal plane a coupling to field-induced antiferroquadrupolar order plays an essential role. The theory was based on a purely localised approach for $4f$ crystalline electric field (CEF) states and it was able to reproduce the qualitative features of the phase diagram. One must keep in mind, however that there is a hybridisation of $4f$ CEF levels with conduction states. This leads to an estimated Kondo temperature of $T^*\simeq 30$ K \cite{hafner:22,christovam:24} which is of the same order as the splitting energy of the lowest two Kramers doublets in the uniaxial CEF potential thus forming a quasi-quartet. On the other hand ARPES experiments suggest that measured band structures reveal primarily localised $4f$ electron character with minor itinerant contributions \cite{chen:24}.\\

In any case it is necessary to map out all physical consequences of the localised CEF state model as a reference point to judge which features are realistic and what may be different in reality due to the presence of Kondo lattice effects. In fact the usefulness of the localised approach for the investigation of multipolar order  and associated phase diagrams has been demonstrated successfully before in other Kondo compounds like \CB\cite{shiina:97,thalmeier:21} and \YR\cite{jeevan:06,takimoto:08}.
The previously developed theory in this spirit for \CR\cite{schmidt:24} focused exclusively on the question of the phase diagram, i.e.,  the structure and appearance of phase boundaries and order parameters and the constraints imposed by the $C_{4v}$ Ce site symmetry on them. In the present work we develop this theory further  and investigate the experimentally most useful thermodynamic quantities in the localised approach, 
We calculate the temperature dependence of coexisting magnetic and quadrupolar order parameters and confirm
that the phase boundaries approached from the ordered regime are identical to the previous results
obtained by approaching from the disordered regime. Furthermore we calculate specific heat, magnetocaloric cooling rate, thermal expansion and magnetostriction. In particular we consider the type of anomalies of these quantities at the H-T phase boundaries that may be used to track them. For this purpose we employ  the simplified quasi-quartet model for the CEF states which was shown to be analytically tractable \cite{schmidt:24}. We also consider the symmetry elastic constant anomalies expected to arise from the magnetoelastic coupling to the field-induced quadrupolar (FIQ) order parameter. The former are directly associated with
the homogeneous static RPA quadrupolar susceptibilities which we calculate in the para- as well as ordered phases.

Another important issue not yet investigated experimentally concerns the low energy magnetic excitation spectrum.
Within the localised approach we determine the dispersive magnetic exciton mode in the disordered regime and under 
presence of coexisting magnetic and quadrupolar order. We show the type of dispersion and intensity to be expected and
predict their temperature and field dependences.

\section{The quasi-quartet model for C\lowercase{e}R\lowercase{h}$_2$A\lowercase{s}$_2$
and the effective operator MF approximation}
\label{sec:quaqua}
%
% %%%%%%%%%%%%%%%%%%%%% figure %%%%%%%%%%%%%%%%%%%%%%%%%%%%
\begin{figure}
\vspace{0.1cm}
\centering
\includegraphics[width=0.70\columnwidth]{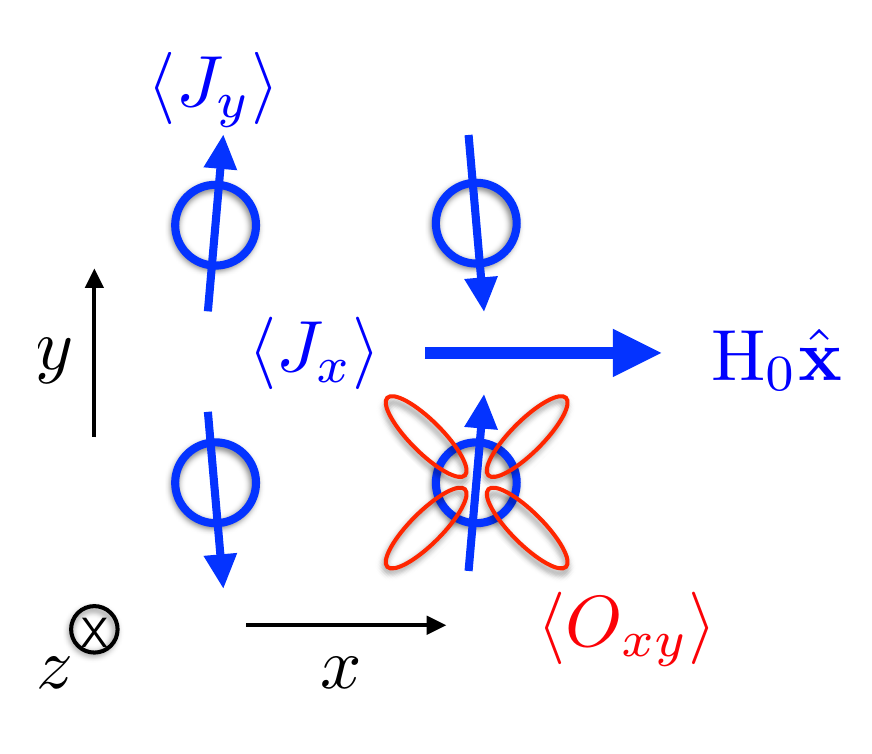}
\caption{Schematic presentation of tetragonal in-plane geometric configurations for the phase diagrams.
The circles show the Ce positions in the plane with the tetragonal a- axes a aligned with x,y coordinates and the
c-axis oriented along z, perpendicular to the plane. Field ${\bf H}_0={\rm H}_0\hat{{\bf x}}$ is along x-direction and low- field moments $\langle J_y\rangle$ oriented along y-direction  and staggered along both x,y directions according to
${\bf Q}=(\pi,\pi)$ ordering vector leading to two AF sublattices $S_\lambda$ $(\lambda=\uparrow,\downarrow \;{\rm or}\; \pm 1)$. For larger field they show canting producing the homogeneous polarization $\langle J_x\rangle$. The field-induced quadrupole order parameter $\langle O_{xy}\rangle$ is also indicated.}
\label{fig:struc}
\end{figure}
%%%%%%%%%%%%%%%%%%%%%%fig%%%%%%%%%%%%%%%%%%%%%%%%%%%%%%%
%
%
% %%%%%%%%%%%%%%%%%%%%% figure %%%%%%%%%%%%%%%%%%%%%%%%%%%%
\begin{figure}
\vspace{0.1cm}
\centering
\includegraphics[width=0.95\columnwidth]{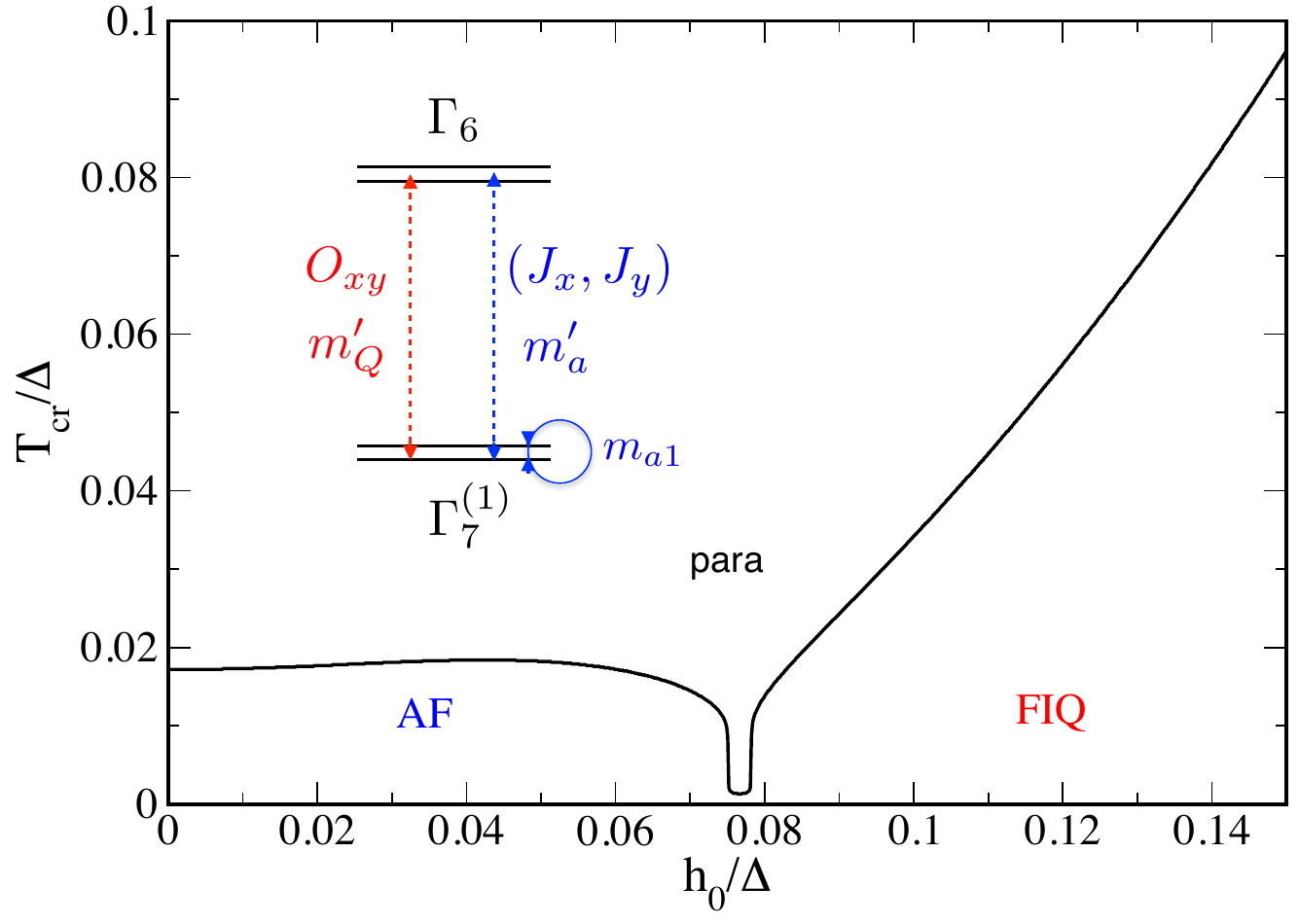}
\caption{Boundaries $T_{cr}(H_0)$ of magnetic (AF) and field-induced quadrupolar (FIQ) phases derived from Fig.~\ref{fig:OPT} tracing the vanishing order parameters. Interaction parameters correspond to QCP endpoint 
with $I_m=0.019$ and $I_Q=0.01166$\cite{schmidt:24} and critical field $h_{0cr}^a/\Delta= 0.077$.
The phase diagram agrees with the one obtained from
the paramagnetic side by tracing the singularity of paramagnetic RPA susceptibilities \cite{schmidt:24}. Inset:
The reduced quasi-quartet system consisting of $\Gamma_7^{(1)}$ and $\Gamma_6$ split by $\Delta=30~\rm K$.
Important matrix elements of dipolar $(m_{a1},m'_a)$ (blue) and quadrupolar $(m'_Q)$ (red) operators are indicated,
for details see Ref.~\cite{schmidt:24}.}
\label{fig:OPphase}
\end{figure}
%%%%%%%%%%%%%%%%%%%%%%fig%%%%%%%%%%%%%%%%%%%%%%%%%%%%%%%
%
An extensive discussion of the CEF model for \CR~and its various simplifications has been presented in Ref.~\cite{schmidt:24} to which we refer for the details. In this section we recapitulate only the minimum necessary
ingredients of this model to make the present extended investigation of its thermodynamic properties self-contained. The  localised Ce$^{3+}$ 4f states with total angular momentum $J=\frac{5}{2}$ split into three Kramers doublets in the tetragonal CEF which are representations of $C_{4v}$ site symmetry (which lacks local  inversion, but this has little influence on the restricted 4f subspace). The ground state doublet $|1\pm\rangle$ is $\Gamma_7^{(1)} (0\;{\rm K})$ and the two excited doublets are  $\Gamma_6 (30\;\text K)$ denoted by  $|2\pm\rangle$  and $\Gamma_7^{(2)} (180\;\text{K})$ termed  $|3\pm\rangle$. For presently accessible (static) applied magnetic fields and temperatures in the range of the ordering temperature we may ignore the upper doublet and restrict to a quasi quartet model, however using its proper tetragonal symmetry wave functions \cite{schmidt:24}.
%
% %%%%%%%%%%%%%%%%%%%%% figure %%%%%%%%%%%%%%%%%%%%%%%%%%%%
\begin{figure}
%\vspace{1cm}
\centering
\includegraphics[width=0.94\columnwidth]{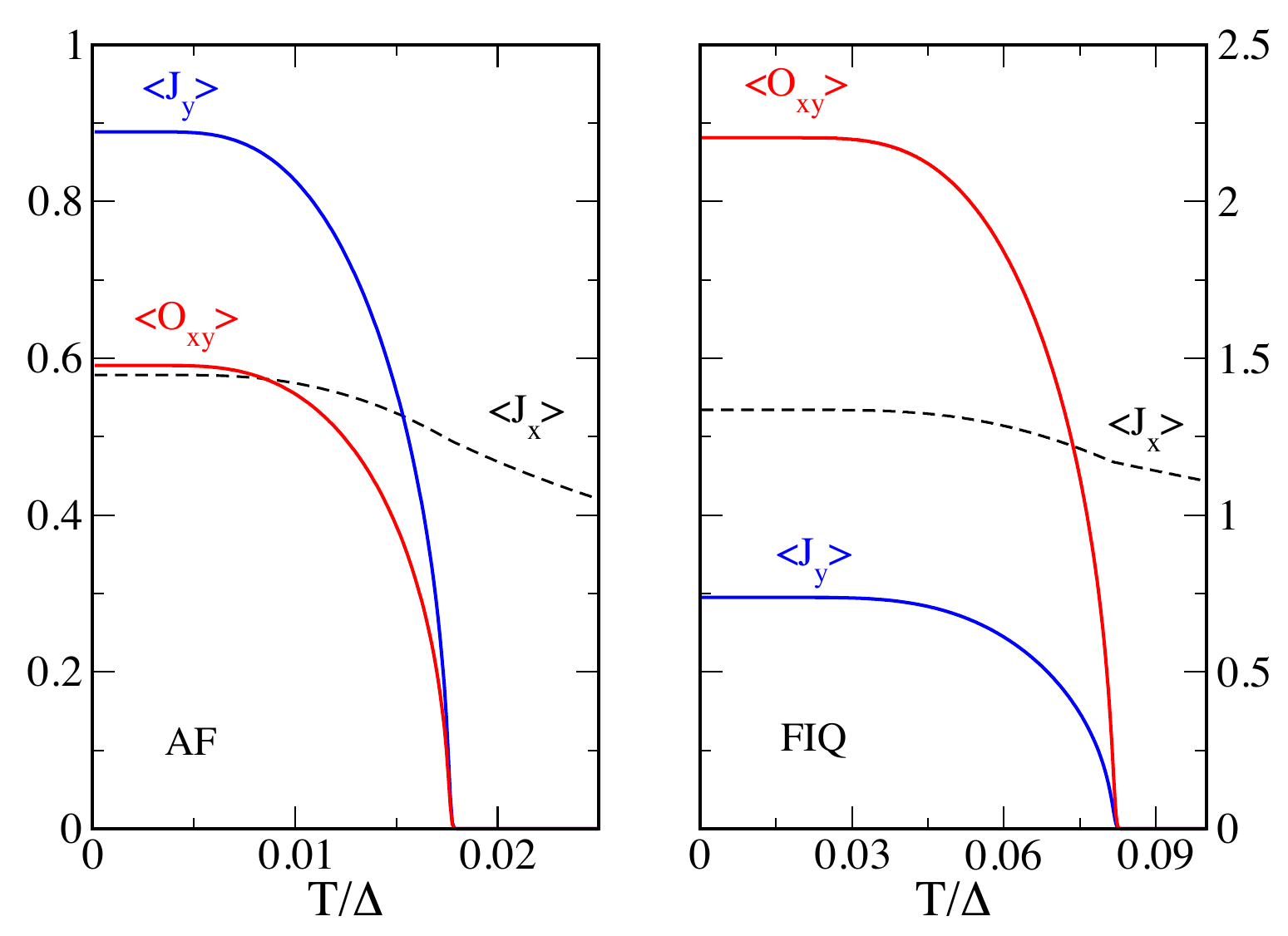}
\caption{Temperature dependence of AF order parameter $\langle J_y\rangle$ and quadrupolar $\langle O_{xy}\rangle$ for 
fields $h_0=0.02\Delta$ deep in AF phase (left panel) and  $h_0=0.14\Delta$ deep in FIQ phase (right panel), together with homogeneous polarisation $\langle J_x \rangle$. In AF panel  $\langle J_y\rangle$ is primary order parameter and $\langle O_{xy}\rangle$ field-induced and vice versa in  the FIQ panel. }
\label{fig:OPT}
\end{figure}
%%%%%%%%%%%%%%%%%%%%%%fig%%%%%%%%%%%%%%%%%%%%%%%%%%%%%%%
%
From high-temperature susceptibility anisotropy obtained directly \cite{hafner:22} (Supp. Mat.) or from NMR results \cite{kitagawa:22} one may expect that a prospective  AF ordered state at zero field has moments oriented within the tetragonal plane (Fig.~\ref{fig:struc}). The most interesting structure of the phase diagram occurs when  the external field ${\bf H}_0$  is also lying in the plane perpendicular to the tetragonal c-axis. Since $(J_x,J_y)$ form a degenerate $\Gamma_5$ doublet the moment can be freely rotated in the tetragonal plane. An applied field will immediately flop the moment directions perpendicular to the field and their canting leads to
a homogeneous polarization in addition to the staggered order parameter perpendicular to the field. We select the field direction along x (the tetragonal a-axis) for simplicity and then $\langle J_y\rangle$ is the dipolar order parameter oriented perpendicular to the field  (Fig.~\ref{fig:struc}).
In this case dipolar $J_y$ moments will be coupled to field induced quadrupolar $O_{xy}$ moments. This situation is described by the Hamiltonian
\begin{eqnarray}
H&=&
-\frac{\Delta}{2}
\sum_i(|1\sigma\rangle\langle1\sigma|-|2\sigma\rangle\langle2\sigma|)_i\nonumber\\
 &&-\frac12\sum_{\langle ij\rangle}J^a_{ij}(J_i^xJ_j^x +J_i^yJ_j^y)
 -\frac12\sum_{\langle ij\rangle} J_{ij}^cJ_i^zJ_j^z\nonumber\\
&&-\frac12\sum_{\langle ij\rangle}J^Q_{ij}O_{xy}(i)O_{xy}(j)-g_J\mu_{\text B}\mu_0\sum_i H_0 J_x
\label{eq:HAM}
\end{eqnarray}
where the first term ($=H_{CF}$) describes the quasi-quartet split by $\Delta=30\; \text{K}$, ($\sigma$ is the Kramers pseudo spin) followed by the effective
magnetic and quadrupolar interaction terms (for in-plane nearest neighbor (n.n.) sites $\langle i,j\rangle$) and the Zeeman term with $h_0=g_J\mu_B\mu_0H_0$.
For a symmetry table of possible couplings of multipole moments see Ref.~\cite{schmidt:24}. It was argued there that for the chosen in-plane geometry (Fig.~\ref{fig:struc}) with field along x-axis (tetragonal a-axis) the $\Gamma_4$- quadrupole $O_{xy}$ is the only possibility to obtain the overall structure of the phase diagram.  In the molecular field (MF) approximation the above Hamiltonian is given by
\begin{eqnarray}
H_{\rm mf}^\lambda(i)&=&H_{CF}(i)-\bigl[h_xJ_x(i)+h^\lambda_yJ_y(i)+h_Q^\lambda O_{xy}(i)\bigr]    
\label{eq:HMF}                               
\end{eqnarray}       
%
%
% %%%%%%%%%%%%%%%%%%%%% figure %%%%%%%%%%%%%%%%%%%%%%%%%%%%
\begin{figure}
\vspace{0.2cm}
\centering
\includegraphics[width=0.95\columnwidth]{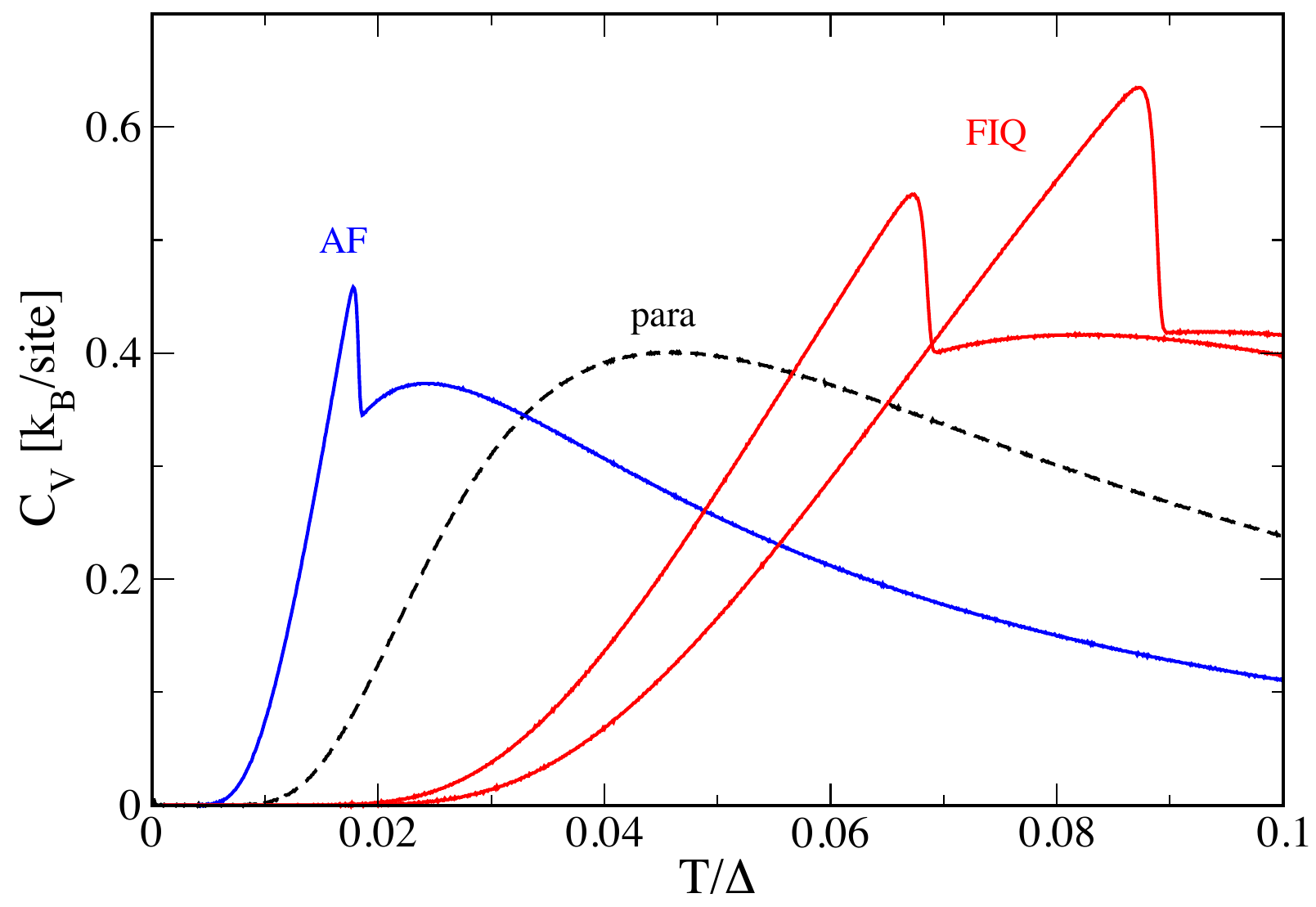}
\caption{Specific heat in the three regions for QCP endpoint parameters of Fig.\ref{fig:OPphase} (touching AF and FIQ phases at $h_0^{cr}\simeq0.08\Delta$). For AF $h_0=0.05\Delta$ and for FIQ $h_0/\Delta=0.13,0.145$ (increasing with $T_Q$). At the critical field the $C_V$ of the para state corresponds to pure Schottky anomaly  of the split ground state doublet.}
\label{fig:CVT}
\end{figure}
%%%%%%%%%%%%%%%%%%%%%%fig%%%%%%%%%%%%%%%%%%%%%%%%%%%%%%%
%
Here we introduced the polarisation or  magnetisation  $\langle J_x\rangle$ (in units of $g_J\mu_B$ per site)  via the homogeneous effective field $h_x=h_0-I_m\langle J_x\rangle$ and the molecular fields associated with the staggered order parameters according to  $h_y^\lambda =\lambda h_y;\; h_y=-I_m\langle J_y\rangle$ and  $h_Q^\lambda =\lambda h_Q;\; h_Q=-I_Q\langle O_{xy}\rangle$ where the two magnetic sublattices $S_\lambda$ correspond to setting $\lambda=\pm 1$ in these expressions. In the restricted quasi-quartet model for $h_0,T\ll \Delta$ it is possible to eliminate  the terms of the Zeeman
and multipolar intersite interaction terms (within MF approximation) that couple the first excited doublet states $|2\pm\rangle$  to the split ground $|1\pm\rangle$ state pair using Brillouin Wigner perturbation theory \cite{zeiger:73}, described in detail in  Ref. \cite{schmidt:24}.
 In this way one retains only an effective split doublet level scheme $|\psi_{n\lambda}\rangle$, $(n=\pm)$  which are the eigenstates of the effective $2\times 2$ Hamiltonian for each sublattice $S_\lambda$:
\begin{eqnarray}
H^\lambda_{eff}&=&
\left(
 \begin{array}{cc}
-\frac{\Delta^*}{2}& \frac{\hat{\delta}_\lambda}{2} \\
\frac{\hat{\delta}^*_\lambda}{2}&-\frac{\Delta^*}{2}\\
\end{array}
\right);\;\;\;
%-\frac{\Delta^*}{2}&=&-\frac{\Delta}{2}-\frac{1}{\Delta}\bigl[{m'}_Q^2h^2_Q+{m'}_a^2(h_x^2+h_y^2)\bigr]\nonumber\\
-\frac{\hat{\delta}_\lambda}{2}=\frac12(\hat{\delta}_1+i\lambda\hat{\delta}_2)\nonumber\\
\Delta^*&=&\Delta+\frac{1}{\Delta}({m'}^2_Qh_Q^2+{m'}_a^2h_m^2)\nonumber\\
\hat{\delta}_1&=&-2\bigl(m_{a1}h_x+\frac{2}{\Delta}m'_am'_Qh_yh_Q\bigr)\nonumber\\
\hat{\delta}_2&=&2\bigl(m_{a1}h_y+\frac{2}{\Delta}m'_am'_Qh_xh_Q\bigr)
\label{eq:Heff}
\end{eqnarray}
The diagonal element $-\Delta^*/2$ is a renormalised level position that plays only a role for the
internal energy and specific heat calculation (Sec.~\ref{sec:thermo}). The eigenvalues including the effect of the polarisation and order parameters
as well as the eliminated upper doublet are then obtained as
\begin{eqnarray}
E_\pm=
%-\frac{\Delta^*}{2}\mp\frac12|\hat{\delta}|=
-\frac{\Delta^*}{2}\mp\frac12(\hat{\delta}_1^2+\hat{\delta}_2^2)^\frac12
\label{eq:splitting}
\end{eqnarray}

And the eigenstates $|\psi_{\pm\lambda}\rangle$ of the effective ground state doublet in terms of the unperturbed CEF ground state $|1\pm\rangle$ are given by the columns of the unitary transformation matrix
\begin{eqnarray}
U_\lambda^\dag=\frac{1}{\sqrt{2}}
\left(
 \begin{array}{cc}
1& e^{i\lambda\phi} \\
-e^{-i\lambda\phi}&1\\
\end{array}
\right);\;\;\;
\tan\phi=\hat{\delta}_2/\hat{\delta}_1
\label{eq:transform}
\end{eqnarray}
Using the matrix $\{W_{n'n}^{\lambda *}\}_{kl}= U^{\lambda *}_{n'k}U^\lambda_{nl}  \equiv    \{W_{nn'}^{\lambda}\}_{lk}$ the 
 effective matrix elements of multipole operators $A$ within the low energy $|\psi_{\pm\lambda}\rangle$ subspace may then
be expressed as 
\begin{eqnarray}
\hat{A}_\lambda\equiv \{A\}^\lambda_{n'n}=
\langle \psi_{n'\lambda}|A\rm|\psi_{n\lambda}\rangle=tr(W^{\lambda *}_{nn'}\cal A^\lambda\rm_{eff})
\label{eq:multimat1}
\end{eqnarray}

For an effective Hermitean $2\times 2$ matrix $A^\lambda_{\rm eff}$ with real diagonal elements $(a_0,a_0)$ and complex nondiagonal ones
$(a_1, a^*_1)$ we obtain
\begin{eqnarray}
\hat{A}_\lambda
&=&
\left(
 \begin{array}{cc}
\hat{a}_{0+}& \hat{a}_1^* \\
\hat{a}_1&\hat{a}_{0-}
\end{array}
\right)
=
\left(
 \begin{array}{cc}
a_0-\tilde{a}_0&\hat{a}'_1-i\hat{a}''_2 \\
\hat{a}'_1+i\hat{a}''_2&a_0+\tilde{a}_0
\end{array}
\right)\nonumber\\[0.3cm]
&=&
\left(
 \begin{array}{cc}
a_0-\text{Re}(e^{i\lambda\phi}a_1^*)& \frac12(a_1^*-e^{-2i\lambda\phi}a_1) \\
 \frac12(a_1-e^{2i\lambda\phi}a^*_1) & a_0+\text{Re}(e^{i\lambda\phi}a_1^*)\\
\end{array}
\right)
\label{eq:multimat2}
\end{eqnarray}
where we have explicitly
\begin{eqnarray}
\tilde{a}_0
&=&(c_{\phi}a'_1+s_{\phi}a''_1)\nonumber\\
\hat{a}'_1&=&(a'_1-c_{2\phi}a'_1-s_{2\phi}a''_1)/2\nonumber\\
\hat{a}''_1&=&(a''_1+c_{2\phi}a''_1-s_{2\phi}a'_1)/2
\label{eq:multimat3}
\end{eqnarray}
%
%
%%%%%%%%%%%%%%%%%%%%%% figure %%%%%%%%%%%%%%%%%%%%%%%%%%%%
\begin{figure}
%\vspace{1cm}
\centering
\includegraphics[width=0.99\columnwidth]{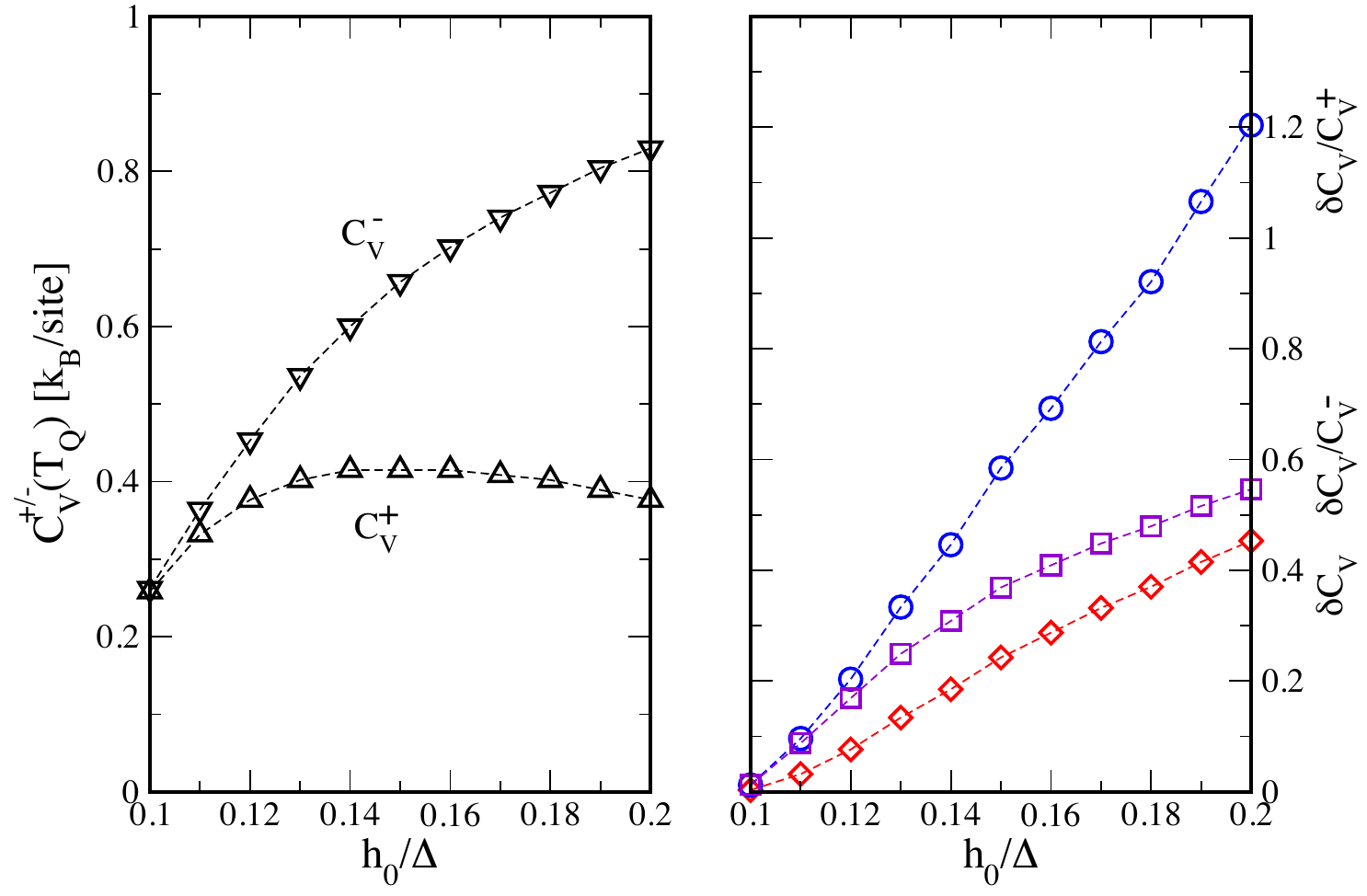}
\caption{Specific heat at transition temperature $T_Q$ in FIQ regime as function of applied field (cf. Fig.\ref{fig:CVT} with same interaction parameters as in Fig.~\ref{fig:OPphase}). Left panel shows $C^\pm_V(T_Q)$ above and below the transition and right panel the absolute and relative step sizes where $\delta C_V=C^-_V-C^+_V$. Lines are guides to the eye. }
\label{fig:CVstep}
\end{figure}
%%%%%%%%%%%%%%%%%%%%%%%%%%%%%%%%%%%%%%%%%%%%%%%%%%%%%%%%%

with $c_\phi=\cos\phi, s_\phi=\sin(\lambda\phi)=\lambda\sin\phi$ and likewise
 $c_{2\phi}=\cos2\phi, s_{2\phi}=\sin(2\lambda\phi)=\lambda\sin2\phi$. The matrix 
 elements  in $\hat{A}_\lambda$ together with the split ground state energies in Eq.~(\ref{eq:splitting})
 determine the multipolar response functions of the quasi quartet model.
 The effective operators $\cal{A}^\lambda_{\text{eff}}$ for the relevant multipoles entering Eq.~(\ref{eq:multimat1}) are given by
 \begin{eqnarray}
J_x^{\text{eff}}
&=&
\left(
 \begin{array}{cc}
\frac{2}{\Delta}{m'}_a^2h_x& m_{a1}-\frac{2i}{\Delta}{m'}_a^2h_Q^\lambda\\
 m_{a1}+\frac{2i}{\Delta}{m'}_a^2h_Q^\lambda&\frac{2}{\Delta}{m'}_a^2h_x
\end{array}
\right)\nonumber\\[0.1cm]
J_y^{\text{eff}}
&=&
\left(
 \begin{array}{cc}
\frac{2}{\Delta}{m'}_a^2h^\lambda_y&-i[m_{a1}+\frac{2i}{\Delta}{m'}_a^2h_Q^\lambda]\\
i[m_{a1}-\frac{2i}{\Delta}{m'}_a^2h_Q^\lambda]&\frac{2}{\Delta}{m'}_a^2h^\lambda_y
\end{array}
\right)\nonumber\\[0.1cm]
O_{xy}^{\text{eff}}
&=&
\left(
 \begin{array}{cc}
\frac{2}{\Delta}{m'}_Q^2h^\lambda_Q& -\frac{2i}{\Delta}{m'}_a {m'}_Q h_+^\lambda\\
 \frac{2i}{\Delta}{m'}_a {m'}_Q h_-^\lambda& \frac{2}{\Delta}{m'}_Q^2h^\lambda_Q
\end{array}
\right)
\label{eq:multeff}
\end{eqnarray}
where $h_\pm^\lambda=h_x\pm ih_y^\lambda$. The various elastic and inelastic
CEF matrix  elements of the bare operators are indicated in the inset of Fig.~\ref{fig:OPphase} (for further details we see  Ref.~\cite{schmidt:24}).
The above effective operators and also the corresponding multipole matrix elements in Eq.~(\ref{eq:multimat3}) contain the influence of the external field, the homogeneous polarisation and both molecular fields. They also incorporate the effect of the eliminated excited $|2\pm\rangle$ doublet states as caused by their non-diagonal couplings $(m_a',m'_Q)$ connecting them to the ground state. These molecular fields can be calculated \cite{schmidt:24} from  selfconsistent MF equations reproduced in Eq.(\ref{eq:appMFA}) as obtained by using the diagonal elements in Eq.~(\ref{eq:multimat1}) for each of the operators $\hat{A}_\lambda$.
With the above multipole operators projected down to only the split ground state doublet the single ion susceptibilities  as well as collective susceptibilities, first static and later dynamical, may readily be obtained. \\
%
% %%%%%%%%%%%%%%%%%%%%% figure %%%%%%%%%%%%%%%%%%%%%%%%%%%%
\begin{figure}
\vspace{0.2cm}
\centering
\includegraphics[width=0.90\columnwidth]{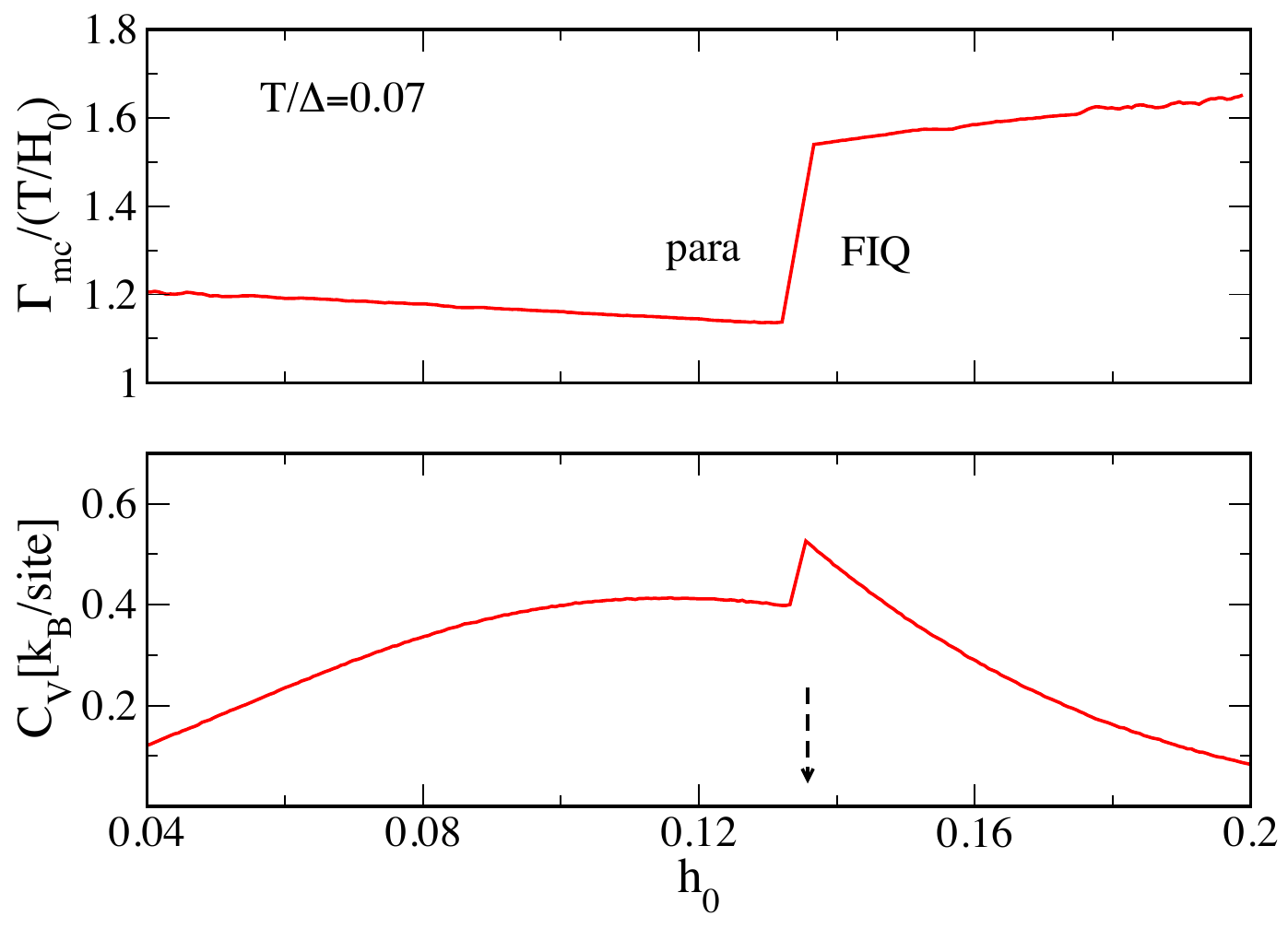}
\caption{Upper panel: Normalized magnetocaloric coefficient (Eq.~(\ref{eq:magcal})) as function of applied field
in a-direction. A step at the para/FIQ phase boundary (critical field $h^a_{0cr}$ marked by arrow) appears concomitantly with the specific heat anomaly (lower panel, cf. Fig.~\ref{fig:CVT})}
\label{fig:caloric}
\end{figure}
%%%%%%%%%%%%%%%%%%%%%%fig%%%%%%%%%%%%%%%%%%%%%%%%%%%%%%%
%

 \section{Thermodynamic potentials and associated anomalies in thermodynamic coefficients}
\label{sec:thermo}

Before proceeding to the response functions we calculate various useful thermodynamic quantities that have been employed to construct the H-T phase diagram by using the observed anomalies upon crossing the boundary.  In particular we consider the specific heat, thermal expansion and magnetostriction and in addition the magnetocaloric coefficient. For this purpose we derive first the necessary thermodynamic potentials explicitly by using the projected effective ground state model derived in the previous section.\\
%
% %%%%%%%%%%%%%%%%%%%%% figure %%%%%%%%%%%%%%%%%%%%%%%%%%%%
\begin{figure}
\vspace{0.2cm}
\centering
\includegraphics[width=0.99\columnwidth]{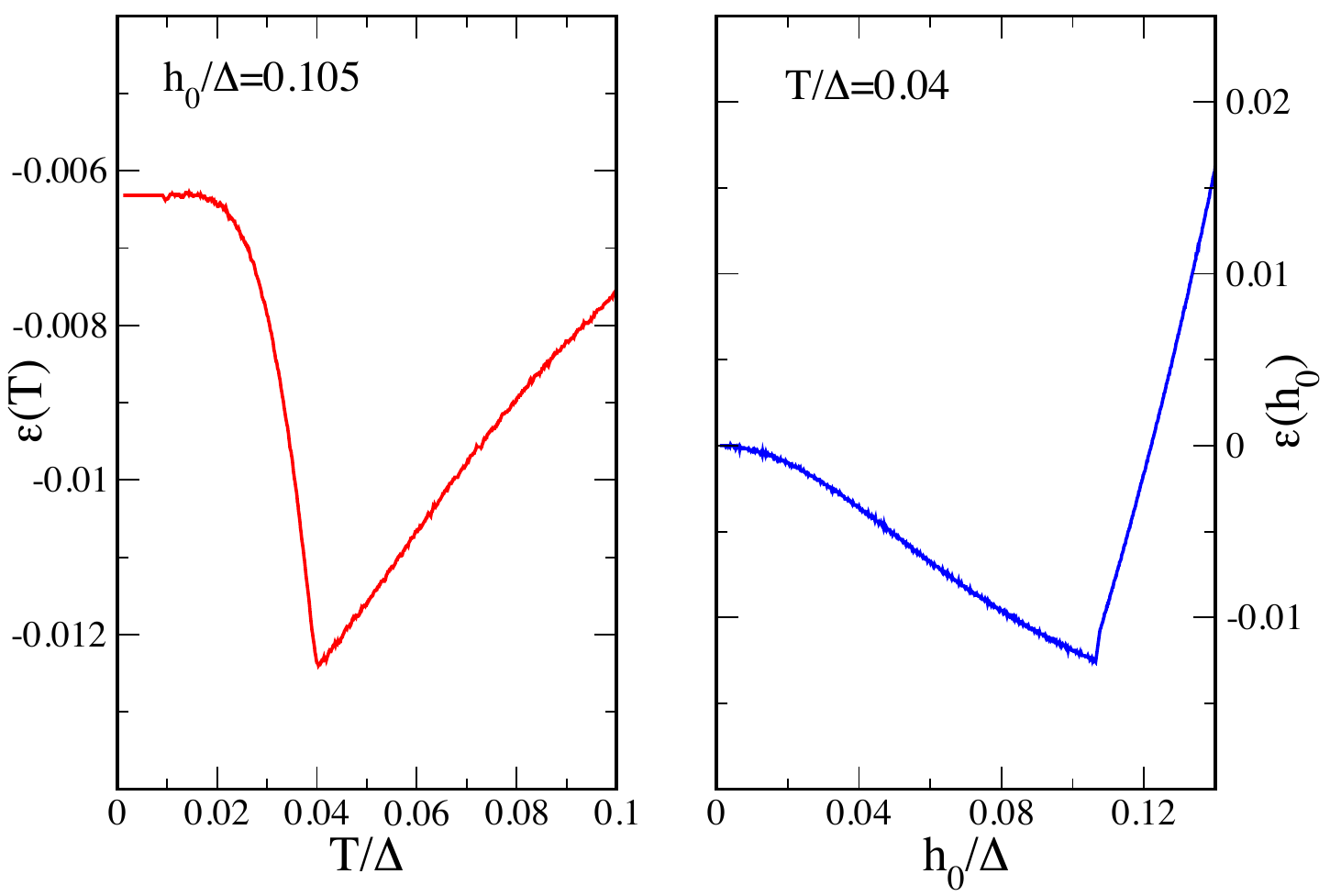}
\caption{left panel: Thermal expansion strain at constant applied field. The kink appears at the critical FIQ temperature
$T_Q(h_0)/\Delta=0.04$. right panel: magnetostriction strain at constant temperature. The kink appears at the critical field
$h^a_{0cr}(T)/\Delta=0.105$ (cf. Fig.~\ref{fig:OPphase}). In both panels the strain unit is $(\hat{g}/c_0)$ (positive or negative depending on the sign of the coupling constant $\hat{g}$).}
\label{fig:striction}
\end{figure}
%%%%%%%%%%%%%%%%%%%%%%fig%%%%%%%%%%%%%%%%%%%%%%%%%%%%%%%
%

{\it i) internal energy and specific heat}\\
First we calculate the total internal energy $U(T,H_0)$ from which the specific heat  $C_V=(\partial U/\partial T)_V$ may be obtained by numerical differentiation. In MF approximation the latter contains the energies of single-ion levels via the selfconsistent ground state splitting as well as the direct order parameter terms. The former is given by $U_0=\sum_n p_nE_n$ where the energy levels $E_n$ $(n=\pm)$ are defined in  Eq.~(\ref{eq:splitting}) and their corresponding thermal occupation is $p_n=\exp(-E_n/T)/Z$ with $Z=\sum_m\exp(-E_m/T)$ denoting the partition function. Note that for these extensive quantities the global shift of the two effective levels by $\Delta^*(T,h_0)$ given in Eq.~(\ref{eq:Heff}) is important, contrary to the calculation of the order parameters (Eq.\ref{eq:appMFA}) where it cancels out. We obtain
\begin{eqnarray}
U(T,H_0)&=& -\frac{\Delta^*}{2} -\frac12|\hat{\delta} |\tanh\frac{|\hat{\delta} |}{2T}\nonumber\\
&&- \frac12 \Bigl[ I_m(\langle J_x\rangle^2-\langle J_y\rangle^2)- I_Q\langle O_{xy}\rangle^2\Bigr]
\label{eq:uint}
\end{eqnarray}
Here the first term originates from the field and temperature dependent shift of ground state level position
caused by the effect of molecular field, the second term is due to the changing thermal populations
of split levels and the molecular field dependence of the splitting. The last term is the direct contribution of
the polarisation and order parameters. Because of the complicated coupled field and temperature dependence of the latter
(Eq.~(\ref{eq:appMFA})) the differentiation to obtain the specific heat can only be carried out numerically.\\

{\it ii) Entropy and magnetocaloric adiabatic cooling rate}\\
In a similar manner the entropy per site (units $k_B\equiv 1$) may be calculated from $S=-\sum_np_n\ln p_n$ which leads to
\begin{eqnarray}
S(T,H_0)=\ln\bigl[2\cosh\frac{|\hat{\delta} |}{2T}\bigr]- \frac{|\hat{\delta} |}{2T}\tanh\frac{|\hat{\delta} |}{2T}
\label{eq:entropy}
\end{eqnarray}
From this the magnetocaloric coefficient 
\begin{eqnarray}
\Gamma_{mc}=\Bigl(\frac{\partial T}{\partial H_0}\Bigr)_S&=&
-\Bigl(\frac{\partial S}{\partial H_0}\Bigr)_T\Bigl(\frac{\partial S}{\partial T}\Bigr)^{-1}_{H_0}\nonumber\\
&=&-\frac{T}{C_V}\Bigl(\frac{\partial M}{\partial T}\Bigr)_{H_0}
\label{eq:magcal}
\end{eqnarray}
is again obtained by numerical differentiation. It is reasonable to normalise it to the value for a paramagnet with fully degenerate spin states where it is given by $\Gamma^0_{mc}=(T/H_0)$.  Since we have calculated already $C_V$ it is most convenient to use the last
form in the above equation with $M=\langle J_x\rangle$ denoting the homogeneous magnetisation. The magnetocaloric coefficient determines the adiabatic temperature change when sweeping the field leading to distinct anomalies on crossing the phase boundaries. Since the high-field phase boundary (Ref.~\cite{schmidt:24}) will have to be obtained by pulsed field experiments under adiabatic conditions $\Gamma_{mc}$ is an experimentally relevant quantity. 
\\

{\it iii) Free energy, thermal expansion and magnetostriction}\\
Finally the free energy $F=U-TS$ may then readily be derived as
\begin{eqnarray}
F(T,H_0)&=& -\frac{\Delta^*}{2} - T\ln\bigl[2\cosh\frac{|\hat{\delta}}{2T}\bigr]\nonumber\\
&&- \frac12 \Bigl[ I_m(\langle J_x\rangle^2-\langle J_y\rangle^2)- I_Q\langle O_{xy}\rangle^2\Bigr]
\label{eq:free-4f}
\end{eqnarray}
From this quantity we may calculate directly the lattice strain caused by thermal expansion and magnetostriction
by adding to this localised $4f$ contribution the classical background elastic energy and assuming a strain dependence of the model parameters. The latter has contributions from two fully symmetric strains \cite{kuwahara:97} which modulate the interaction parameters $I_m/\Delta$ and $I_Q/\Delta$ of the Hamiltonian. This would entail four magnetostrictive coupling parameters. Here, for demonstration of the principal effects we simplify this model using only the planar strain $\epsilon=\frac12(\epsilon_{xx}+\epsilon_{yy})$ which is justified since we
consider only in-plane n.n.  interactions. The treatment for the three dimensional case is briefly discussed in Appendix \ref{sec:appstrain}.

Because the (scaled) interaction constants $I_{m,Q}/\Delta$ depend on the strain via the exchange-striction $(I_{m,Q}(\epsilon))$ and 
magnetoelastic $(\Delta(\epsilon))$ effects the free energy of the 4f subsystem also depends on the strain. Together with the classical background elastic energy $F_{el}(\epsilon)=\frac12 c_0\epsilon^2$ ($c_0$ is the background fully symmetric elastic constant) this will lead to new temperature- and field- dependent equilibrium strain 
\begin{eqnarray}
\epsilon(T,h_0)=-\frac{1}{c_0}\Bigl(\frac{\partial F}{\partial \epsilon}\Bigr)_{T,h_0}
\label{eq:striction}
\end{eqnarray}
which gives the thermal expansion (at constant $h_0$) or magnetostriction (at constant $T$) of the coupled system. To compute the
derivative from Eq.~(\ref{eq:striction}) one has to introduce the dimensionless striction constants $\hat{g}_{m,Q}=(I_{m,Q}/\Delta)'/(I_{m,Q}/\Delta)\equiv \hat{g}$ where the prime indicates the derivative with respect to the strain $\epsilon$. For simplicity we assume their equality since both are planar n.n. couplings. Then we have for the strain dependent interactions $(I_{m,Q}/\Delta)_\epsilon=
(I_{m,Q}/\Delta)_0(1+\hat{g}\epsilon)$. Inserting this into the $4f$- part of the free energy (Eq.(\ref{eq:free-4f})) the derivative $(\partial F(\epsilon)/\partial \epsilon)_0$ may be computed numerically which defines the thermal expansion and magnetostriction in Eq.~(\ref{eq:striction}). Thereby the strain dependence of the three molecular fields (Eq.~(\ref{eq:appMFA})) has to be taken into account.
In Sec.~\ref{sec:discussion} we will give a discussion of the results for the thermodynamic quantities $C_V, \Gamma_m$ and $\epsilon$ derived above.

\section{Single-ion susceptibilities in the para phase and ordered regime }
\label{sec:singleion}

In this section we want to determine the important symmetry elastic constant anomalies that
are a direct signature of the homogeneous order parameter (quadrupolar) susceptibility.
We start with the expression for the dynamical single ion response function for an arbitrary CEF system and any Hermitean 
operators $X_\alpha,X_\beta$ acting in this space which is given by the RPA formula \cite{jensen:91}
\begin{eqnarray}
&&\chi^0_{\alpha,\beta}(T,{\bf h},{\rm i}\omega_n)=\nonumber\\
&&\sum_{n\neq m}\langle n|X_\alpha |m\rangle\langle m|X_\beta |n\rangle\frac{p_n-p_m}{E_m-E_n-{\rm i}\omega_n}\nonumber\\
+&&\frac{1}{T}\Bigl[\sum_n\langle n|X_\alpha |n\rangle\langle n|X_\beta |n\rangle p_n-\langle X_\alpha\rangle\langle X_\beta\rangle\Bigr]\delta_{\omega_n,0}
\label{eq:baresus}
\end{eqnarray}
where we assumed presence of an arbitrary small field such that all degeneracies are lifted. In the zero
field limit the correct form of $\chi^0_{\alpha\beta}$ incorporating the degeneracies is automatically recovered. The
sum may run over all CEF states in principle ($n,m=$ $1$..$6$). Mostly such single ion susceptibilities 
are calculated numerically for the bare CEF states in the disordered phase. Here we will also derive them for the more difficult case under the presence of three molecular fields that mix the bare CEF states in a complicated manner  and hence modify the multipolar matrix elements in the expression above.\\

In this investigation we assume that the calculation of dipolar magnetic, quadrupolar and (elastic)
response functions should be carried out analytically as far as possible in order to shed light on the basic origin
and behavior of their physical anomalies at the phase boundaries and inside the ordered regime. 
For this purpose we restrict to the quasi-quartet
model  of Ce$^{3+}$ 4f states outlined above using the elimination procedure for the first excited doublet at $\Delta$ as described before. Then the multipole operators for $A,B=J_x,J_y,O_{xy}$ are projected to their effective forms
by Eqs.~(\ref{eq:multimat1},\ref{eq:multeff}) in the lowest split doublet subspace ($|\psi_{n\lambda}\rangle,n=\pm$).
%
%%%%%%%%%%%%%%%%%%%%%% figure %%%%%%%%%%%%%%%%%%%%%%%%%%%%
\begin{figure}
%\vspace{1cm}
\centering
\includegraphics[width=0.99\columnwidth]{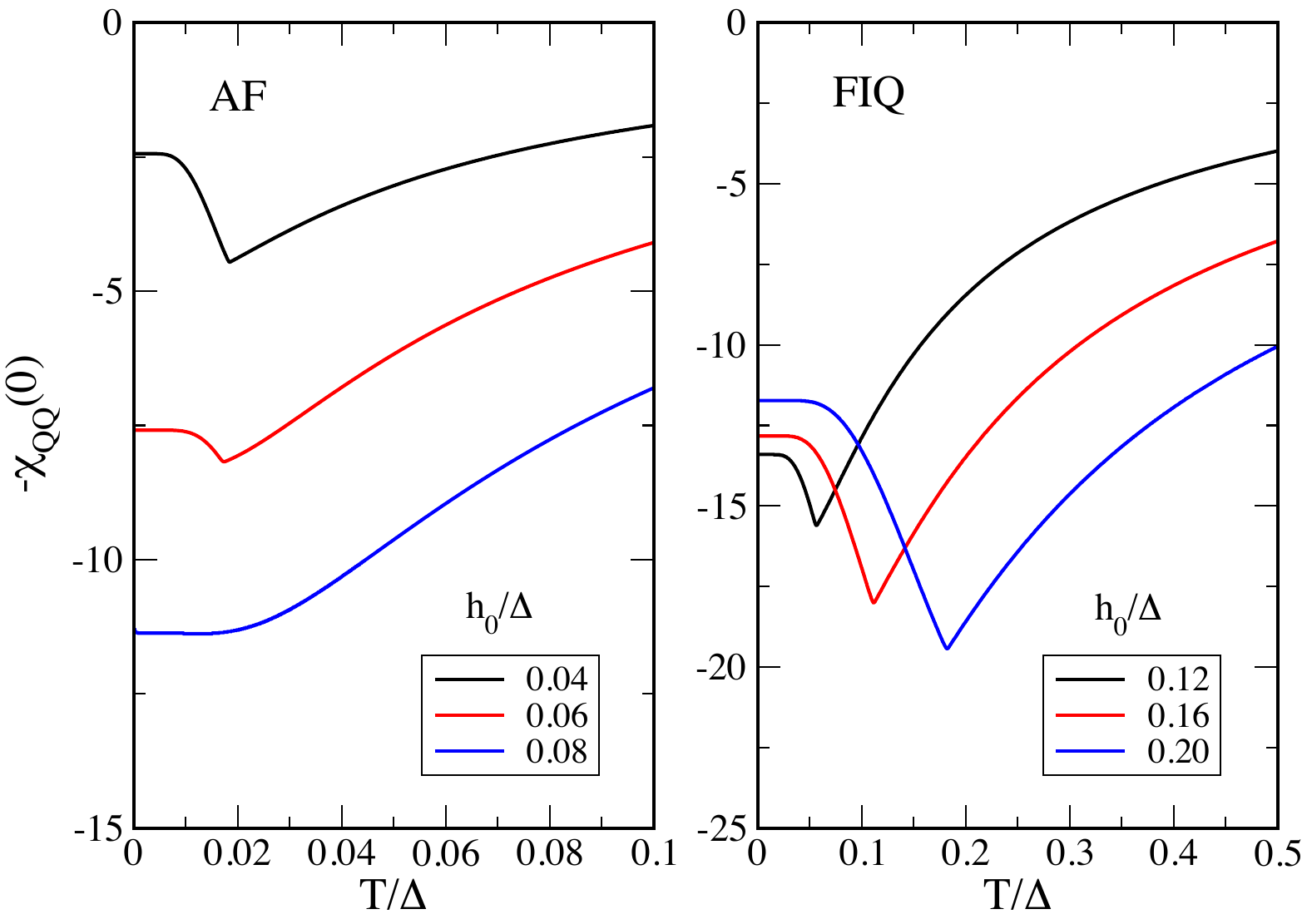}
\caption{The (negative) static homogeneous quadrupolar susceptibility proportional to elastic constant 
anomaly for $c_{66}$ mode (Eq.~(\ref{eq:elastcon})) as function of temperature. Left panel in AF phase. Right panel
for FIQ case, note the different scales for both axes. The anomaly position moves with T$_{cr}(h_0)$
and its size becomes more pronounced in the FIQ regime with increasing $h_0$ (cf. Fig.~\ref{fig:c66-cont}).}
\label{fig:elastcon}
\end{figure}
%%%%%%%%%%%%%%%%%%%%%%fig%%%%%%%%%%%%%%%%%%%%%%%%%%%%%%%
%

\section{Collective quadrupolar susceptibility and elastic constant anomalies}
\label{sec:elast}

The elastic constant anomalies resulting from the magnetoelastic coupling to CEF states
 are related to the homogeneous quadrupolar zero frequency response as described below. 
 Therefore we first need the static single ion susceptibility for $A,B=J_y,O_{xy}$ for para- as well as ordered phase. In the effective
 operator approach for the lowest doublet it is given by 
\begin{eqnarray}
&&\chi^0_{AB}(0)=
\sum_{n\neq m}\langle \psi_n|A_{\rm eff}|\psi_m\rangle\langle \psi_m|B_{\rm eff} |\psi_n\rangle\frac{p_n-p_m}{E_m-E_n}\nonumber\\
+&&\frac{1}{T}\Bigl[\sum_n\langle \psi_n|A_{\rm eff} |\psi_n\rangle\langle \psi_n|B_{\rm eff} |\psi_n\rangle p_n-\langle A_{\rm eff}\rangle\langle B_{\rm eff}\rangle\Bigr]
\label{eq:baresusCVV}
\end{eqnarray}
The first and second terms are vanVleck and pseudo-Curie terms, respectively.
 We note that the single ion susceptibilities do not depend on the sublattice $\lambda$ because $E_n, p_n$ are independent of $\lambda$  and in the combinations it always appears as $\lambda^2=1$, therefore the sublattice label $\lambda$ will be suppressed in the following. Explicitly the three essential bare susceptibilities are given by the general expression
 \begin{eqnarray}
 \chi^0_{AB}(0)&=&\chi^0_{BA}(0)\nonumber\\
 &=&2\text{Re}(\hat{a}_1^*\hat{b}_1)\frac{\tanh\frac{|\hat{\delta}|}{2T}}{|\hat{\delta}|}+
 \frac{1}{T}\frac{\tilde{a}_0\tilde{b}_0}{\cosh^2\frac{|\hat{\delta}|}{2T}}
 \label{eq:susex0}
 \end{eqnarray}
 % 
  %
%%%%%%%%%%%%%%%%%%%%%% figure %%%%%%%%%%%%%%%%%%%%%%%%%%%%
\begin{figure}
\vspace{-0.5cm}
\centering
\includegraphics[width=0.95\columnwidth]{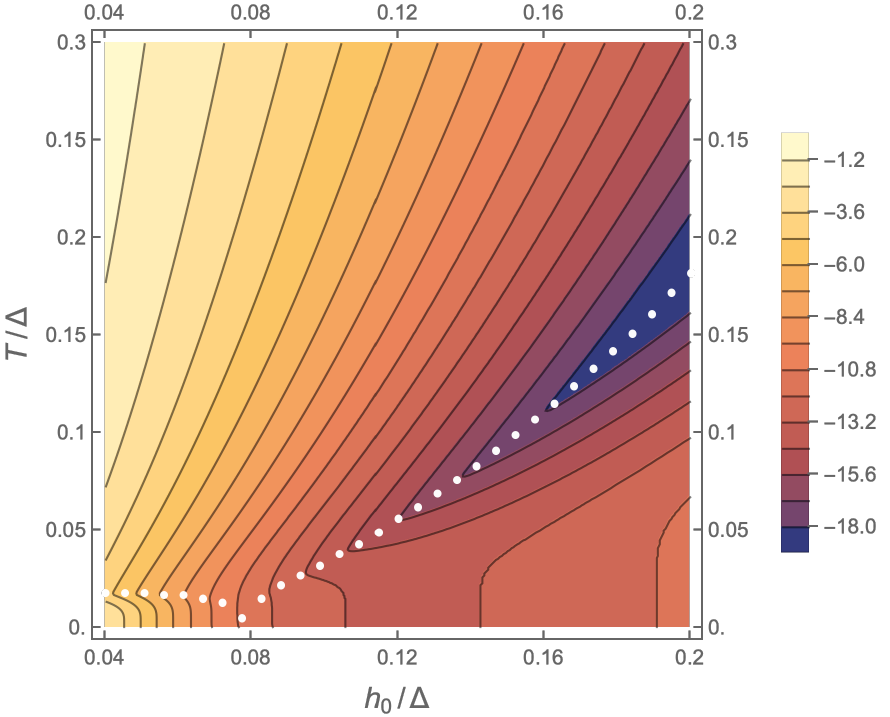}
\caption{Contour plot of $c_{66}$ elastic constant anomaly in the T-h$_0$ plane according to Eq.~(\ref{eq:elastcon})
with $\delta c_{66}/c^0_{66}=-(\tilde{g}^2_{66}/c^0_{66})\chi_{QQ}(T,h_0)$ in units of the dimensionless coupling prefactor.
Following the kinks (white dots) of contours tracks the phase boundaries for
AF phase ($h_0/\Delta<0.08)$ and FIQ phase  ($h_0/\Delta>0.08)$. In the latter the anomalies
become progressively more pronounced for increasing field (cf. Fig.~\ref{fig:elastcon}).  }
\label{fig:c66-cont}
\end{figure}
%%%%%%%%%%%%%%%%%%%%%%fig%%%%%%%%%%%%%%%%%%%%%%%%%%%%%%%
%
where again first and second terms are vanVleck and pseudo-Curie contributions, respectively.
The AB/BA symmetry holds due to the $\text{Re}(\hat{a}_1^*\hat{b}_1)=\text{Re}(\hat{b}_1^*\hat{a}_1)$ symmetry of matrix elements where $(\tilde{a}_0,\hat{a}_1)$ are defined in Eq.~(\ref{eq:multimat3}) and completely analogous relations hold for the pair
 $(\tilde{b}_0,\hat{b}_1)$ for the B multipolar operator.  To calculate the collective quadrupole response functions \cite{schmidt:24} one needs the single ion susceptibilities for the pairs $(A,A)=(J_y,J_y)$, $(B,B)=(O_{xy},O_{xy})$ and $(A,B)=(J_y,O_{xy})$. The latter is nonvanishing for finite field because of field-induced quadrupole matrix elements \cite{schmidt:24}. Then we may define the weight factors appearing in Eq.~(\ref{eq:susex0}) as  $w_a=|\hat{a}_1|^2$,  $w_b=|\hat{b}_1|^2 $ and
 $w_{(a,b)}=\text{Re}(\hat{a}_1^*\hat{b}_1)$ (symmetric in a,b). We note that from this   the relation
$w_{(a,b)}^2=w_aw_b$ holds, so that it is sufficient to calculate only $w_a, w_b$..
 Together with $\tilde{a}_0$ and $\tilde{b}_0$ they determine  the three single ion susceptibilities. Explicitly from Eq.~(\ref{eq:multimat3}) (and equivalent one for the B multipole) we obtain the following expressions in terms of bare CEF matrix elements and molecular fields including  $A=J_x,J_y$ (quantities denoted by a) and $B=O_{xy}$ (quantities marked by b). For the pseudo-Curie terms we have:
\begin{eqnarray}
\tilde{a}_0(J_x)&=&c_\phi m_{a1}-s_\phi(\frac{2}{\Delta}m'_am'_Qh_Q)\nonumber\\
\tilde{a}_0(J_y)&=&c_\phi (\frac{2}{\Delta}m'_am'_Qh_Q)-s_\phi m_{a1} \nonumber\\
\tilde{b}_0(O_{xy})&=&c_\phi(\frac{2}{\Delta}m'_am'_Qh_y)-s_\phi(\frac{2}{\Delta}m'_am'_Qh_x)
\end{eqnarray}
 and likewise for the vanVleck terms we get
 \begin{eqnarray}
  w_a(J_x)&=&\frac12\Bigl[(1-c_{2\phi})m_{a1}^2\nonumber\\
 &&+(1+c_{2\phi})(\frac{2}{\Delta}m'_am'_Qh_Q)^2\nonumber\\
&&+2s_{2\phi}(\frac{2}{\Delta}m_{a1}m'_am'_Qh_Q)\Bigr]\nonumber\\[0.2cm]
 w_a(J_y)&=&\frac12\Bigl[(1-c_{2\phi})(\frac{2}{\Delta}m'_am'_Qh_Q)^2\nonumber\\
 &&+(1+c_{2\phi})m_{a1}^2\nonumber\\
&&+2s_{2\phi}(\frac{2}{\Delta}m_{a1}m'_am'_Qh_Q)\Bigr]\nonumber\\[0.2cm]
w_b(O_{xy})&=&\frac12\Bigl[(1-c_{2\phi})(\frac{2}{\Delta}m'_am'_Qh_y)^2\nonumber\\
 &&+(1+c_{2\phi})(\frac{2}{\Delta}m'_am'_Qh_x)^2\nonumber\\
&&+2s_{2\phi}(\frac{2}{\Delta}m'_am'_Q)^2h_xh_y\Bigr]
%\nonumber\\[0.2cm]
%w_{(a,b)}&=&\frac12\Bigl\{(1-c_{2\phi})(\frac{2}{\Delta}m'_am'_Q)^2h_yh_Q\nonumber\\
% &&+(1+c_{2\phi})(\frac{2}{\Delta}m_{a1}m'_am'_Qh_x)\nonumber\\
%&&+s_{2\phi}\bigl[(\frac{2}{\Delta}m'_am'_Q)^2h_xh_Q
%+\frac{2}{\Delta}m_{a1}m'_am'_Qh_y)\bigr]
%\Bigr\}
\label{eq:susex1}
 \end{eqnarray}
Inserting these weights into the pseudo Curie and vanVleck terms of 
Eq.~(\ref{eq:susex0}) and using the ground state doublet splitting $|\hat{\delta}|$ 
from Eq.~(\ref{eq:Heff}) completes the calculation of the single ion susceptibility
in the ordered phase and under the presence of external and molecular fields.\\

The quantity probed in elastic constant experiments, however, is the collective
quadrupolar RPA susceptibility at zero wave vector and frequency. It contains
the Fourier transforms $I_A({\bf k}), I_B({\bf k})$ of effective intersite interactions of multipoles $A,B$. For the n.n. interaction models \cite{schmidt:24}  we have $I_A(0)=-I_m$ and $I_B(0)=-I_Q$, i.e. simply a sign change with respect to the (positive)
values at the AF zone boundary vector ${\bf k}=(\pi,\pi)$. This is important because it means that close to the phase boundary
of the AF and FIQ phases the staggered RPA susceptibilities (at the zone boundary vector) diverge but not
the homogeneous one (for ${\bf k}=0$) which, as shown below couples to the elastic constant anomalies. The latter are
therefore finite and exhibit no pronounced softening behavior as would be the case for an induced ferroquadrupolar order which  is the case in the quasi-quartet system \YR~\cite{rosenberg:19,takimoto:08}. With this in mind the static homogeneous quadrupolar RPA susceptibility is given by (${\bf k}$ and $\omega=0$ suppressed):
\begin{eqnarray}
\chi_{QQ}&=&\frac
{\chi^0_{QQ}(1+I_m\chi^0_{yy})+I_Q\chi^0_{Qy}\chi^0_{yQ}}
{1+I_m\chi^0_{yy}+I_Q\chi^0_{QQ}+I_mI_Q(\chi^0_{yy}\chi^0_{QQ}-\chi^0_{Qy}\chi^0_{yQ})}\nonumber\\
\label{eq:statQQ}
\end{eqnarray}
This response function determines the elastic anomalies at the phase boundaries and in the ordered phase
probed by ultrasonic velocity measurements. We note that the mixed dipolar-quadrupolar term $\sim I_mI_Q$
can be simplified because the pure vanVleck terms vanish due to the relation
$w_aw_b=w_{(a,b)}^2$ in Eq.~(\ref{eq:susex1}) and likewise the pure pseudo-Curie terms cancel leaving only
the contribution of cross-terms according to 
\begin{eqnarray}
&&\chi^0_{yy}\chi^0_{QQ}-\chi^0_{Qy}\chi^0_{yQ}=\nonumber\\
&& \frac{1}{T}\frac{2}{\cosh^2\frac{|\hat{\delta}|}{2T}}
\frac{\tanh\frac{|\hat{\delta}|}{2T}}{|\hat{\delta}|}
|\hat{a}_1\tilde{b}_0-\hat{b}_1\tilde{a}_0|^2
\label{eq:crossterm}
\end{eqnarray}
This combination appears only for the {\it static } response, for finite frequency according to the RPA expression
there are only vanVleck terms and therefore the mixed dipolar-quadrupolar  terms vanish in the denominator
of Eq.~(\ref{eq:dynyy}) which will simplify greatly the calculation of collective magnetic excitations in the subsequent section.\\

To establish the correlation between elastic anomalies and quadrupolar response one
has to add the magnetoelastic Hamiltonian \cite{thalmeier:91} containing the appropriate elastic strains to the model of Eq.~(\ref{eq:HAM}). Since the nondegenerate $O_{xy}$ quadrupole couples linearly only to the equal symmetry ($\Gamma_4$ of C$_{4v}$) strain  representation  $\epsilon_{xy}$ corresponding to the background elastic constant $c^0_{66}$, then the magnetoelastic Hamiltonian is simply given by
\begin{eqnarray}
H_{me}=\frac12 c^0_{66}\epsilon^2_{xy}-g_{66}\sum_i \epsilon_{xy}O_{xy}(i)
\end{eqnarray}
This leads to a field- and temperature dependent renormalisation of the elastic constant $c_{66}(T,H_0)$  (as compared to $c^0_{66}$) given by the downward correction \cite{thalmeier:91} ($\tilde{g}^2_{66}=Ng^2_{66}$):
\begin{eqnarray}
\hspace{-0.3cm} 
\delta c_{66}(T,H_0)= c_{66}(T,H_0)-c^0_{66}= -\tilde{g}^2_{66}\chi_{QQ}(T,H_0)
\label{eq:elastcon}
\end{eqnarray}
which is thus directly proportional to the static, homogeneous RPA quadrupolar susceptibility. The relative change
$\delta c_{66}(T,H_0)/c^0_{66}$ is controlled by the dimensionless coupling strength $\tilde{g}^2_{66}/c^0_{66}$ which is commonly of the order $10^{-4}..10^{-6}$ for heavy fermion compounds.
Since we have calculated the quadrupolar response function (Eq.~(\ref{eq:statQQ})) for the whole temperature and field range $(T,H_0 \ll \Delta)$
we can predict the elastic constant behavior above as well as below the phase boundaries.
The predicted $\delta c_{66}$ are shown in Figs.~\ref{fig:elastcon},\ref{fig:c66-cont} and discussed further in Sec.~\ref{sec:discussion}.

\section{Low energy excitation spectrum in the AF and FIQ ordered phases}
\label{sec:dynamic}

Another important probe of ordered phases is the dynamical magnetic response 
determined by inelastic neutron scattering. The cross section is proportional
to the magnetic spectral function. We will consider the response both longitudinal
($\parallel $$y$) and transverse ($\parallel$$x$)  direction, refering always  to the AF ordered moment direction $\parallel$$ y$ (Fig.~\ref{fig:struc}).
(Note that in the paramagnetic state the polarised homogeneous moment is $\parallel$ $ x$ and hence the
meaning of longitudinal  and transverse  is reversed). Similar to the static case in Eq.~(\ref{eq:statQQ}) the dynamic 
magnetic response for general wave vector in the BZ
for $A=J_x$ or $A=J_y$ with $B=O_{xy}$ may be written as \cite{jensen:91}
\begin{eqnarray}
\label{eq:dynyy}
&&\chi_{AA}({\bf k},{\rm i}\omega_n)=\nonumber\\
&&\frac
{\chi^0_{AA}({\rm i}\omega_n)[1-I_B({\bf k})\chi^0_{BB}({\rm i}\omega_n)]-I_A({\bf k})\chi^0_{AB}({\rm i}\omega_n)\chi^0_{BA}({\rm i}\omega_n)}
{1-I_A({\bf k})\chi^0_{AA}({\rm i}\omega_n)-I_B({\bf k})\chi^0_{BB}({\rm i}\omega_n)}\nonumber\\
\end{eqnarray}
where we have used the fact that the mixed dipolar-quadrupolar terms in the denominator vanish for finite frequency as explained below Eq.~(\ref{eq:crossterm}).
Here $I_A({\bf k})=-I_m\gamma({\bf k})$ and $I_B({\bf k})=-I_Q\gamma({\bf k})$ where $\gamma({\bf k})=\frac12(\cos k_x +\cos k_y)$ is the n.n. in-plane structure function of the tetragonal lattice.\\ 

The collective magnetic excitation mode resulting from the split (by external and molecular fields)  doublet state is then given by looking at the pole $\omega({\bf k})$  of the above response function on the real axis (${\rm i}\omega\rightarrow \omega+{\rm i}\eta$). For that purpose we first need the dynamical single-ion susceptibilities which are given by
\begin{eqnarray}
\chi^0_{AA}(\omega)&=&\hat{w}_a(T)F(\omega);\
\chi^0_{BB}(\omega)=\hat{w}_b(T)F(\omega)\nonumber\\
\chi^0_{AB}(\omega)&=&\hat{w}_{(a,b)}(T)F(\omega)
\end{eqnarray}
where the abbreviations $\hat{w}(T)=w\tanh(|\hat{\delta} |/2T)$ are used and the weights $w_a,w_b$ and $w_{(a,b)}$ are 
given in Eq.~(\ref{eq:susex1})
The dynamics is contained in ($\tilde{\eta}_\omega=2\eta\omega$)
\begin{eqnarray}
F(\omega)=\frac{2|\hat{\delta} |}{|\hat{\delta}|^2-{\rm i}\omega_n^2-{\rm i}\tilde{\eta}_\omega}
\end{eqnarray}
Then the poles of the magnetic response function lead to the collective magnetic mode dispersion as given by
\begin{eqnarray}
\omega_{\bf k}^2=|\hat{\delta} |\bigl[|\hat{\delta} | +2\gamma_{\bf k}(I_mw_a+I_Qw_b)\tanh\frac{|\hat{\delta}|}{2T}\bigr]
\end{eqnarray}
which is formally the same expression for longitudinal $(A=J_y)$ and transversal $(A=J_x)$ response provided
that the proper $w_a(J_y)$ and  $w_a(J_x)$ matrix element combinations (Eq.~(\ref{eq:susex1})) in each case are used.
%
%%%%%%%%%%%%%%%%%%%%%% figure %%%%%%%%%%%%%%%%%%%%%%%%%%%%
\begin{figure}
%\vspace{1cm}
\centering
\includegraphics[width=0.99\columnwidth]{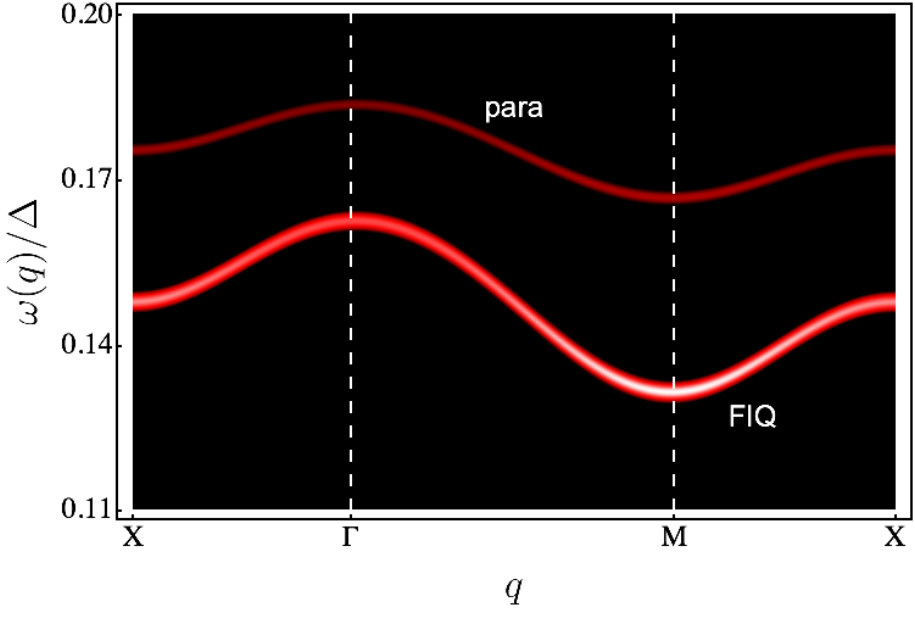}
\caption{Intensity plot of the $yy$ dynamical response function (Eq.~(\ref{eq:spectrum})) in the para and FIQ phases  corresponding to two values of $(h_0,T/\Delta)$. Lower dispersion:  FIQ- phase $(0.1,0.01)$ (dashed arrow in Fig.~\ref{fig:Mpoint}). 
Upper dispersion:  para- phase $(0.11,0.15)$. In the former the energy, dispersion and intensity (proportional to brightness) increase with respect to the para phase. The Intensity maximum of the FIQ dispersion is at the M-point and the minimum at $\Gamma$. Interaction parameters as in Fig.~\ref{fig:OPphase}. A broadening of the spectral function with $\eta=0.002$ was used. The in-plane tetragonal BZ points are (units of $\pi/a$):
$X(0,1), \Gamma(0,0), M(1,1)$.}
\label{fig:specyy}
\end{figure}
%%%%%%%%%%%%%%%%%%%%%%fig%%%%%%%%%%%%%%%%%%%%%%%%%%%%%%%
%
The field- and temperature dependent intensity of the collective magnetic modes relevant for INS (without the Bose factor) is obtained from the spectrum of Eq.~(\ref{eq:dynyy}) via
\begin{eqnarray}
&&\hat{{\cal S}}_a({\bf k},\omega)=\frac{1}{\pi}Im\chi_{AA}({\bf k},\omega+i\eta)_{\eta\rightarrow 0}\nonumber\\
&&=\Bigl(\frac{|\hat{\delta} |}{\omega_{\bf k}}\Bigr)
\Bigl(\frac{w_aI_m(w_a-w_b)}{w_aI_m+w_bI_Q}\Bigr)\tanh\frac{|\hat{\delta}|}{2T}\delta(\omega-\omega_{\bf k})
\label{eq:spectrum}
\end{eqnarray}
It is useful to consider a few limiting cases for the longitudinal $(yy)$ mode  to obtain some qualitative insight. The transverse $(xx)$
mode has small dispersion and is of little interest in the high-field regime.\\

\begin{figure}
\vspace{0.0cm}
\centering
\includegraphics[width=0.95\columnwidth]{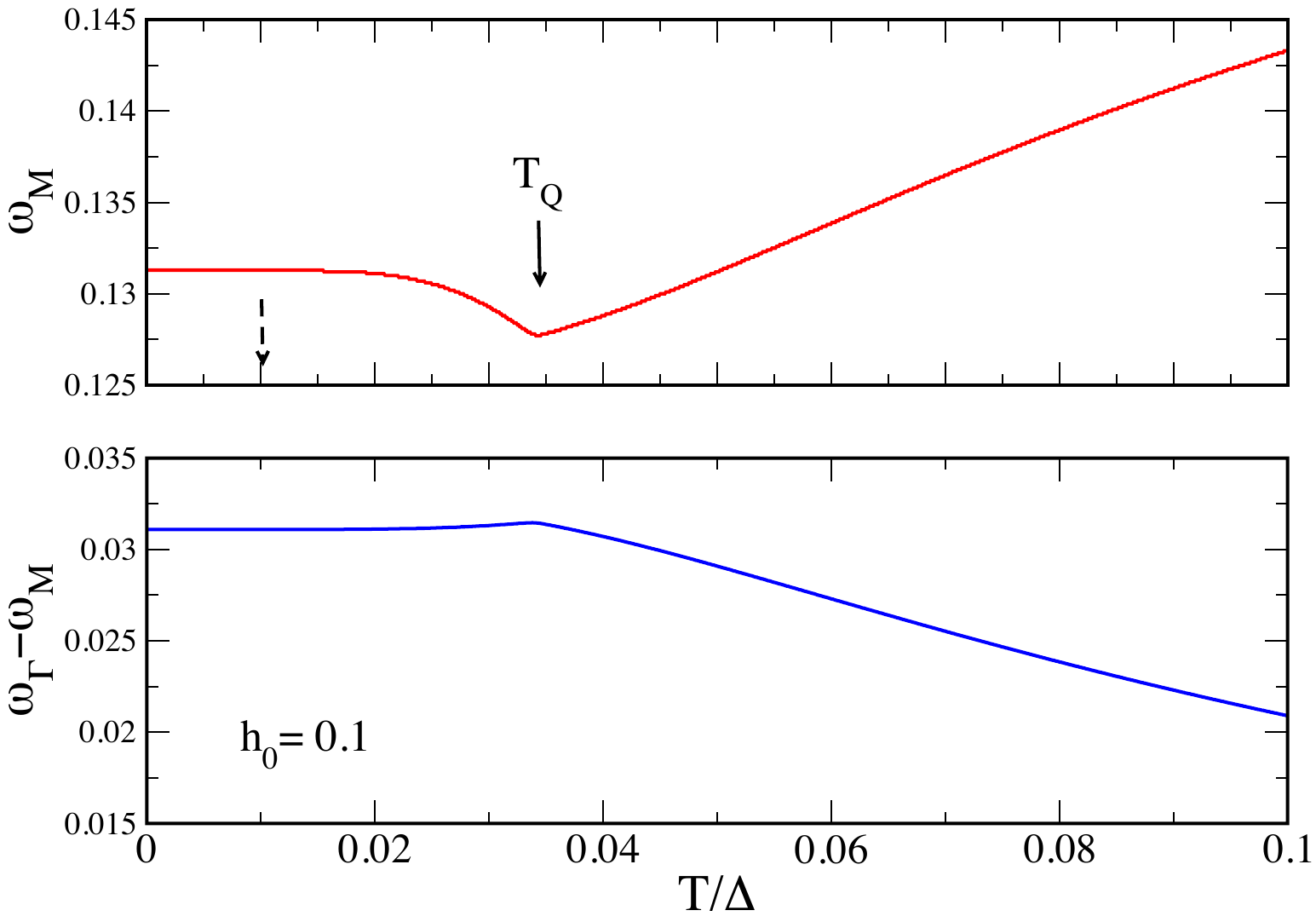}
\caption{Top panel: Temperature dependence of low-energy, high-intensity M-point longitudinal mode in 
Fig.~\ref{fig:specyy} for $h_0=0.1$. The softening for $T>T_Q$ is reversed into rehardening below $T_Q$ due to appearance of magnetic and quadrupolar molecular fields. Arrow corresponds to lower curve in Fig.~\ref{fig:specyy}. Bottom panel: In contrast the overall dispersive width $\omega_\Gamma-\omega_M$ in Fig.~\ref{fig:specyy} is almost constant in the ordered regime but decreases for $T>T_Q$ due to thermal population effects.}
\label{fig:Mpoint}
\end{figure}

{\it i) The paramagnetic phase at finite fields}\\
In this case $h_y=h_Q=0$ and $h_x=h_0-I_m\langle J_x\rangle$. With the resulting simplified weights $w_a, w_b$ we get
\begin{eqnarray}
\omega^2_{\bf k}&=&|\hat{\delta}|\Bigl(|\hat{\delta}|+2I_e\gamma_{\bf k}\tanh\frac{|\hat{\delta}|}{2T}\Bigr)\nonumber\\
I_e&=&m^2_{a1}I_m+(\frac{2}{\Delta}m'_am_Qh_x)^2I_Q
\end{eqnarray}
where $I_e$ is an effective paramagnetic interaction constant that contains the influence of virtual excitations to the first
excited doublet. When temperature increases at fixed $h_0$ 
the dispersion is gradually reduced (see Figs.~\ref{fig:specyy},\ref{fig:Mpoint}) leaving only the local excitation energy $|\hat{\delta}|$ for $T\gg|\hat{\delta}|$.\\

{\it ii) The FIQ phase at large fields and low temperatures}\\
In this limit the population factor approaches unity and $|\hat{\delta}|$ becomes large compared to the
interaction energy scales. Then we get
\begin{eqnarray}
\omega_{\bf k}=|\hat{\delta}|+\gamma_{\bf k}(I_mw_a+I_Qw_b)
\end{eqnarray}
where the general expressions in Eq.~(\ref{eq:susex1}) have to be used since all
molecular fields are nonvanishing. The behaviour of longitudinal mode dispersion and
intensities are discussed further in Sec.~\ref{sec:discussion}.
%
%%%%%%%%%%%%%%%%%%%%%% figure %%%%%%%%%%%%%%%%%%%%%%%%%%%%
%%%%%%%%%%%%%%%%%%%%%%fig%%%%%%%%%%%%%%%%%%%%%%%%%%%%%%%
% 

\section{Discussion of thermodynamics, elastic constants and excitation spectrum in the quasi quartet model}
\label{sec:discussion}

Now we discuss the physical effects that have been derived in the previous sections for the quasi-quartet 
model of \CR, their most interesting aspect being the behavior of anomalies at the phase boundaries.

The latter have been
investigated in detail in a previous work \cite{schmidt:24} by approaching the phase boundary
from the disordered side and tracing the singularity of the collective RPA susceptibilities. Here,
the results of a complementary approach from the ordered regime are shown in Figs.\ref{fig:OPphase},\ref{fig:OPT}. By 
tracing the critical temperature of the vanishing order parameters from the AF or FIQ phase
the phase diagram may also be constructed. It agrees with the analytical
calculation from the paramagnetic side \cite{schmidt:24} but is less accurate due to the necessary numerical
evaluation of  orders parameter close to the phase boundary. The inset of Fig.~\ref{fig:OPphase} depicts the quasi-quartet level scheme and its essential multipole matrix elements.

The specific heat behavior (Sec.~\ref{sec:thermo}) as function of temperature is shown in Fig.~\ref{fig:CVT} for various applied fields 
in the AF (blue) and FIQ (red) regime. Right at the critical field $h^a_{0cr}$ the para phase is recovered and
the specific heat is of Schottky-type due to the split ground state doublet. Away from the critical
field the specific heat jumps at the phase boundary are superposed on the background Schottky anomaly.

The evolutions of the specific values at temperatures slightly above and below the jump and the jump size itself are presented in Fig.~\ref{fig:CVstep} for the FIQ phase. In the low field regime the absolute as well as relative
 step sizes increase with field, reflecting the increase of primary quadrupolar and 
 induced magnetic order parameters.

The complementary field dependence of specific heat at constant temperature is shown in Fig.~\ref{fig:caloric} (lower panel)
which likewise exhibits a jump anomaly when crossing into the FIQ phase. Its appearance is inverted as compared
to Fig.~\ref{fig:CVstep} because the FIQ phase is reached from the para phase for decreasing temperature but increasing 
field. In the top panel the normalised magnetocaloric coefficient $\Gamma_{mc}/\Gamma^0_{mc}$ (Eq.~(\ref{eq:magcal}))
is also shown as function of applied field. It exhibits a jump anomaly at the critical field. The adiabatic temperature change
$\Delta T_{ad}$ within a finite field interval is the integral over $\Gamma_{mc}$. Therefore if the interval sweeps the critical field, $\Delta T_{ad}$ should show a slope change at the critical field.

The related quantities of thermal expansion and magnetostriction are depicted in Fig.~\ref{fig:striction} as temperature and field sweeps through the same point $(T,h_0)=(0.04,0.105)$ on the phase boundary (see Fig.~\ref{fig:OPphase}) on the FIQ side, therefore the cusp minima have the same value. Their sharpness makes both quantities a preferred method for tracking the phase boundary. At low temperature  the thermal expansion approaches the zero temperature value exponentially while the magnetostriction tends to zero quadratically with field strength.

A most direct method to probe quadrupolar ordering is the investigation of symmetry elastic constants $c_\Gamma(T,h_0)$ \cite{thalmeier:91} (Sec.~\ref{sec:elast}).  If the symmetry corresponds to the one of the order parameter then it probes directly the homogenous order parameter susceptibility according to Eq.(\ref{eq:elastcon}) (i.e. at wave vector ${\bf k}\simeq 0$). For a ferroquadrupolar transition this would result in a divergence of the latter and concomitant elastic constant softening as a precursor to a structural transition. In \CR~ we are dealing with the antiferroquadrupolar case for $O_{xy}$ symmetry with ${\bf k}=(\pi,\pi)$ and hence $\chi_{QQ}(T,h_0)$ will not diverge, likewise the associated $c_{66}(T,h_0)$ symmetry elastic constant will not soften but show a kink-like appearance like the
thermodynamic quantities. Its evolution as function of field is shown in Fig.~\ref{fig:elastcon} with the small kink due to induced quadrupole in the AF gradually vanishing at the critical field (left panel) and then strongly recovering in the FIQ phase due to primary quadrupole order parameter. This evolution of $c_{66}$ elastic constant anomaly is comprehensively shown in Fig.~\ref{fig:c66-cont}
as a contour plot showing the deepening valley of the anomaly for higher fields in the FIQ phase. Following the evolution of the valley  (white dots) traces perfectly the phase boundary of Fig.~\ref{fig:OPphase}. This method should also be applicable in the high field regime of the phase diagram accessible only with pulsed fields.

Finally we come to the results for dynamical properties represented by the  spectrum of magnetic excitations (Sec.~\ref{sec:dynamic}).
We focus here on the higher field FIQ phase where the modifications due to Kondo effect may be expected to play a less prominent role.
The Fig.~\ref{fig:specyy} shows the dispersive spectrum associated with the $(yy)$ dynamical response function (Eq.~(\ref{eq:spectrum})). For higher temperatures in the para phase the dispersion is relatively small and intensity weak, both effects caused by the smaller level population difference at high temperatures. The center of gravity corresponds to the level splitting $\hat{\delta}$ at the given field $h_0$. For low temperatures within the FIQ phase the dispersion is larger and the intensity has a maximum at the ordering wave vector (M-point). Note the fields in the two cases are slightly different. For the FIQ dispersion the temperature evolution of the M-point frequency is shown in Fig.~\ref{fig:Mpoint}. It (upper panel) exhibits a softening in the paramagnetic phase to a minimum value when crossing the FIQ phase boundary ($T_Q(h_0)$) and a subsequent rehardening at lower temperatures. The overall dispersion width as function of temperature is shown in the lower panel. We finally note that a complete softening of magnetic excitations at the transition cannot occur since at a finite field on the FIQ side induces a strong quadrupole which breaks the in-plane continuous rotation symmetry of the AF order parameter so that no corresponding Goldstone mode can be expected.

\section{Summary and Conclusion}
\label{sec:summary}

In this work we investigated physical effects in a localised 4f- model that is relevant for the 
new heavy fermion superconductor \CR. The coupling of itinerant electrons and Kondo physics
is not included in this model. It would require to start from a microscopic quasi-quartet Kondo lattice model
such as has been proposed for Yb- compounds in Ref.~\cite{thalmeier:18, akbari:20}. However, it was shown in a previous work \cite{schmidt:24} that nevertheless the localised 4f model is able to explain the basic H-T phase diagram of the normal state broken symmetry phases in \CR, in particular its prominent a-c anisotropy.  It has been argued there that competing antiferromagnetic
and field-induced antiferroquadrupolar order is at the origin of the anomalous phase diagram.\\

Here we have calculated in detail the most relevant characteristics of  thermodynamics, elastic response and
magnetic excitation spectrum in this model. These calculations have been carried out within a simplified 
quasi-quartet model which allows a mostly analytical treatment. We showed that the phase diagram
obtained by approaching the phase boundary from the broken symmetry side agrees well with
the results previously obtained from the paramagnetic side. The specific heat anomalies and their 
evolution with field has been mapped in the FIQ phase. Likewise the adiabatic magnetocaloric
cooling rate exhibits a steplike discontinuity at the phase boundary. This property should be quite useful
in future pulsed high field experiments which are performed under adiabatic conditions.
Furthermore the thermal expansion and magnetostriction show sharp dip-like anomalies which 
for this reason are well suited to track the phase boundary. 
A most direct way to probe the FIQ phase is the determination of $c_{66}$ symmetry elastic constants which
are influenced only by he FIQ order of $O_{xy}$ type. Cusp-like anomalies are predicted that grow in the 
high field regime. This is a signature of the FIQ order parameter and should be investigated experimentally.
Furthermore we have deduced the magnetic excitation spectrum and its field and temperature evolution
in the ordered phase. In particular the high-field case may be amenable and useful for experiments on single crystal \CR~using inelastic neutron scattering.\\

\appendix

%\section{Bare multipole operators for CEF states}
%\label{sec:appCEF}

\section{Selfconsistent MF equations}
\label{sec:appMF}

Using the effective operator representations in Eq.~(\ref{eq:multeff}) the MF expectation values in the case $T,h_0\ll\Delta$ may finally be calculated from  $\langle A\rangle_\lambda=\sum_{n=\pm}p_n\langle\psi_{n\lambda}|A|\psi_{n\lambda}\rangle$ with the help of Eq.~(\ref{eq:multimat1}). Explicitly this leads to the coupled equations for the homogeneous magnetisation $\langle J_x\rangle$ and staggered order parameters $\langle J_y\rangle_\lambda, \langle O_{xy}\rangle_\lambda$ (for AF sublattice $S_\lambda$). They were derived in Ref. \cite{schmidt:24} where the origin and meaning of individual terms is explained:
\begin{eqnarray}
\langle J_x\rangle=&&\bigl[-m_{a1}\cos\phi+\frac{2}{\Delta}m'_am'_Qh_Q\sin\phi\bigr]\tanh\frac{|\hat{\delta}|}{2T}
+\frac{2}{\Delta}{m'_a}^2h_x \nonumber\\
\langle J_y\rangle_\lambda=-\lambda&&\bigl[-m_{a1}\sin\phi+\frac{2}{\Delta}m'_am'_Qh_Q\cos\phi\bigr]\tanh\frac{|\hat{\delta}|}{2T}
+\lambda\frac{2}{\Delta}{m'_a}^2h_y \nonumber\\
\langle O_{xy}\rangle_\lambda=\lambda&&\bigl[\frac{2}{\Delta}m'_am'_Q(h_x\sin\phi-h_y\cos\phi)\bigr]\tanh\frac{|\hat{\delta}|}{2T}
+\lambda\frac{2}{\Delta}{m'_Q}^2h_Q
\label{eq:appMFA}
\end{eqnarray}
The mixing angle $\phi$ is defined in Eqs.~(\ref{eq:Heff},\ref{eq:transform}).
The numerical solution of these coupled equations gives the field- and temperature dependence of homogeneous magnetisation and
order parameters. The former was shown in Ref.~\cite{schmidt:24} an example of the latter is presented in Fig.~\ref{fig:OPT} in this work. By following the temperature where the order parameters $\langle J_y\rangle$ and $\langle O_{xy}\rangle$ vanish the phase boundary in Fig.~\ref{fig:OPphase} may be constructed.

\section{Volume strain and uniaxial strain for the C$_{4v}$ symmetry}
\label{sec:appstrain}

Here we give the general three dimensional form of the thermal expansion and magnetostriction relation in Eq.~(\ref{eq:striction})
used for the two dimensional  in-plane case.
In three dimensional $D_{4h}$ symmetry there are two fully symmetric ($\Gamma_1)$ strains \cite{kuwahara:97}:
The volume strain 
$\epsilon_{\alpha 1}=(1/\sqrt{3})(\epsilon_{xx}+\epsilon_{yy}+\epsilon_{zz})$ and the uniaxial strain
$\epsilon_{\alpha 2}=\sqrt{2/3}[\epsilon_{zz}-\frac12(\epsilon_{xx}+\epsilon_{yy})]$.
Their contribution to the elastic energy is given by
\begin{eqnarray}
F_{el}=\frac12 c_{\alpha 1}\epsilon^2_{\alpha 1} + \frac12 c_{\alpha 2}\epsilon^2_{\alpha 2} + \frac12 c_{\alpha 12}\epsilon_{\alpha 1}\epsilon_{\alpha 2}
\label{eq:free-el}
\end{eqnarray}
where the symmetry elastic constants $c_{\alpha i}$ and $c_{\alpha 12}$ are given in Ref.~\cite{kuwahara:97}. The other relevant 
strain variable in this work is the trigonal $\Gamma_4$ in-plane strain $\epsilon_{xy}$ which couples to the quadrupolar order parameter in first order leading to the $c_{66}$ elastic constant anomaly described in Sec.\ref{sec:thermo}. Adding $F_{el}$ to the 4f- electron part Eq.~(\ref{eq:free-4f}) and looking for the minimum the new equilibrium strains (which preserve the lattice symmetry) are obtained as
\begin{eqnarray}
\epsilon_{\alpha 1}&=&\frac{1}{C_{d}}\Bigl[c_{\alpha 12}\bigl(\frac{\partial F}{\partial\epsilon_{\alpha 2}}\bigr) -
c_{\alpha 2}\bigl(\frac{\partial F}{\partial\epsilon_{\alpha 1}}\bigr)\Bigr]\rightarrow 
-\frac{1}{c_{\alpha 1}}\bigl(\frac{\partial F}{\partial\epsilon_{\alpha 1}}\bigr)\nonumber\\
\epsilon_{\alpha 1}&=&\frac{1}{C_{d}}\Bigl[c_{\alpha 12}\bigl(\frac{\partial F}{\partial\epsilon_{\alpha 1}}\bigr) -
c_{\alpha 1}\bigl(\frac{\partial F}{\partial\epsilon_{\alpha 2}}\bigr)\Bigr]\rightarrow 
-\frac{1}{c_{\alpha 2}}\bigl(\frac{\partial F}{\partial\epsilon_{\alpha 2}}\bigr)\nonumber\\
\end{eqnarray}
with $C_d=c_{\alpha 1}c_{\alpha 2}-c_{\alpha 12}^2$. The limits indicated by arrows correspond to $c_{\alpha 12}\ll (c_{\alpha 1},c_{\alpha 2})^\frac12$
and are each identical in form to the in-plane case of Eq.~(\ref{eq:striction}), but with a doubling of the number
of exchange-striction parameters.\\

\section*{Acknowledgments}
The authors thank M. Brando, E. Hassinger, S. Khim and S. Galeski for helpful discussion on experimental issues.

%%%%%%%%%%%%%%%%%%%%%%%%      References        %%%%%%%%%%%%%%%%%%%%
%\newpage

\section*{References}

\bibliographystyle{iopart-num}
\bibliography{References}
\end{document}